\documentclass{aa}

\usepackage{graphicx}
\usepackage{txfonts}
\usepackage{lipsum}
\usepackage{subcaption}         
\usepackage{lscape}             
\usepackage{placeins}           
\usepackage{xcolor}
\usepackage{wasysym}

\def\chisqr{\hbox{$\chi^2_{\rm r}$}}
\def\msun{\hbox{${\rm M}_{\odot}$}}
\def\mjup{\hbox{${\rm M}_{\jupiter}$}}
\def\me{\hbox{${\rm M}_{\oplus}$}}

\def\msw{\hbox{$\dot {\rm M}_{\odot}$}}
\def\rsun{\hbox{${\rm R}_{\odot}$}}
\def\lsun{\hbox{${\rm L}_{\odot}$}}

\def\mstar{\hbox{$M_{\star}$}}
\def\rstar{\hbox{$R_{\star}$}}
\def\lstar{\hbox{$L_{\star}$}}
\def\teff{\hbox{$T_{\rm eff}$}}
\def\logg{\hbox{$\log g$}}

\def\vD{\hbox{$v_{\rm D}$}}

\def\ms{\hbox{m\,s$^{-1}$}}

\def\kms{\hbox{km\,s$^{-1}$}}
\def\cmss{\hbox{cm\,s$^{-2}$}}

\def\vsini{\hbox{$v \sin i$}}

\def\mic{\hbox{$\mu$m}}

\def\emr{}
\def\Bl{\hbox{$B_{\rm \ell}$}}
\def\Bd{\hbox{$B_{\rm d}$}}

\def\degr{\hbox{$^\circ$}}

\def\Prot{\hbox{$P_{\rm rot}$}}

\newcommand{\hei}{\hbox{He$\;${\sc i}}}

\newcommand{\pab}{\hbox{Pa${\beta}$}}

\begin{document}


\title{Searching for close-in planets around TWA~7 with SPIRou} 

   \author{J.-F.~Donati\inst{1}
      \and P.I.~Cristofari\inst{2}
      \and C.~Moutou\inst{1}
      \and A.~Lavail\inst{1}
      \and J.~Bouvier\inst{3}
      \and S.H.P.~Alencar\inst{4} 
      \and P.~Petit\inst{1}
      \and A.~Carmona\inst{1} 
      \and X.~Delfosse\inst{3} 
      \and the SPIRou science team
          }
   \institute{Univ.\ de Toulouse, CNRS, IRAP, 14 avenue Belin, 31400 Toulouse, France\\ \email{jean-francois.donati@irap.omp.eu}
         \and Leiden Observatory, Leiden University, Niels Bohrweg 2, 2333 CA Leiden, the Netherlands
	 \and Univ.\ Grenoble Alpes, CNRS, IPAG, 38000 Grenoble, France 
         \and Departamento de F\'{\i}sica -- ICEx -- UFMG, Av. Ant\^onio Carlos, 6627, 30270-901 Belo Horizonte, MG, Brazil
             }

\date{Submitted 2025 September -- Accepted 2025 xxx} 

\abstract{
We outline in this paper observations of the young pre-main-sequence low-mass star TWA~7, hosting a debris disk and a distant planet.  Using data collected with the 
near-infrared SPIRou spectropolarimeter / precision velocimeter at the Canada-France-Hawaii Telescope from early 2019 to mid 2021, we detected the magnetic field of 
TWA~7 from the circularly polarized Zeeman signatures and Zeeman broadening of atomic spectral lines, and the rotational modulation of its longitudinal 
component at the known stellar rotation period (of $5.012\pm0.007$~d).  We then modeled the large-scale and small-scale magnetic properties of TWA~7 using Zeeman-Doppler 
imaging.  We found that TWA~7 hosted a mainly poloidal field that significantly evolved from 2019 to 2021, the dipole component getting stronger ({\emr increasing from 0.5~kG in 
2019 to} 0.7~kG in 2021) and less inclined to the stellar rotation axis ({\emr from 22\degr\ in 2019 to} 15\degr\ in 2021).  We also analyzed the radial velocities of TWA~7 derived 
from the SPIRou data, and found a tentative planet signature at a period of 15.2~d (with aliases at 20.8 and 30.4~d), very close to the detection limit of our data and that 
would correspond to a 0.17~\mjup\ planet at a distance of 0.09~au if confirmed.  We finally report modulation of the 1083~nm \hei\ and 1282~nm \pab\ lines of TWA~7 with a 
period of 6.6~d, different from the rotation period and potentially hinting at the presence of a close-in planet triggering star-planet interactions.  
} 


\maketitle



\section{Introduction}
\label{sec:int}

Tremendous observational and theoretical efforts have been invested over the last few decades on the study of planetary systems of low-mass stars, and how they form and 
evolve over their life time \citep[e.g.,][]{Drazkowska23}.  During the initial pre-main-sequence (PMS) stage, both the central host star and its future planets form more 
or less at the same time from a giant molecular cloud that collapses under the combined effect of gravity, turbulence and magnetic fields \citep{Andre14,Lebreuilly24}.  
This early phase of planet formation is still relatively unconstrained on the observational side.  It reflects the extremely challenging task of detecting young planets 
around PMS stars that exhibit a huge level of photometric and spectroscopic variability caused by multiple phenomena occurring at the heart of the accretion disk, including 
stochastic accretion, eruptive events and rotational modulation \citep{Cody14,Sousa16}.  Magnetic fields play a key role throughout this formation process \citep{Suin25}, 
including in the core regions of the accretion disk where the host star generates a magnetospheric gap, with the field funnelling the disk material inward to the stellar 
surface and outward through outflows, and controlling the stellar angular momentum until most of the disk gas and dust are exhausted \citep{Zanni13,Bouvier14,Romanova21}.  
Magnetic fields can also potentially stop inward migrating planets at the outer edge of the magnetospheric cavity \citep{Lin96,Romanova19}, thereby creating pile ups of close-in 
bodies \citep{Romanova06,Mulders15} and saving them from falling into their host stars.  

As a result of the extreme variability of these young stars, the most successful techniques at detecting extra-solar planets, namely the transit method based on 
continuous photometry (yielding estimates of planetary radii) and the radial velocity (RV) technique collecting data from ultra-stable spectrographs (yielding planetary 
masses) struggle to achieve reliable detections for a large sample of targets younger than 20~Myr.  The few confirmed detections so far were obtained on the least active 
low-mass PMS stars, among which {\emr are} AU~Mic {\emr \citep[][aged $\simeq$20~Myr]{Martioli21,Donati25}} and V1298~Tau {\emr \citep[][aged $\simeq$10~Myr]{David19,Finociety23}} 
with their close-in multi-planet systems, and IRAS~04125+2902, the youngest of all {\emr \citep[][aged $\simeq$3~Myr]{Barber24b}}, featuring at least one close-in planet.  
While some of these planets seem to be born with a bulk density comparable to that of the icy giants of the 
Solar System, some others, and in particular those closest to their host stars, have much lower densities, indicating puffed-up radii or a different formation process 
\citep{Finociety23b,Barat24,Donati25b,Donati25c}.  More observations are obviously needed before a statistically significant sample of close-in planets younger than 20~Myr 
is available.  

Among the known young planets, a few of the most massive were detected with direct imaging, at distances of tens of au from their host stars, like the two planets around 
PDS~70 \citep{Haffert19} and the one recently found in the debris disk of TWA~7 \citep{Lagrange25}, a 10~Myr star located in the young TW~Hydra association (TWA).  Searching 
for close-in planets around these stars can bring key constraints on planet formation models by providing information on the architectures of young planetary systems, rather than 
on only one side of the planet distribution with {\emr orbital distance}.  No close-in planet was detected around PDS~70 from velocimetric observations, with an upper limit on 
the mass of such planets of $\simeq$4~\mjup\ at a distance of $\simeq$0.2~au \citep[corresponding to a period of $\simeq$30~d;][]{Donati24}.  

In the present study, we analysed spectropolarimetric 
and velocimetric data of TWA~7 collected with the SPIRou near-infrared (nIR) spectropolarimeter \citep{Donati20} at the 3.6-m Canada-France-Hawaii Telescope (CFHT) over three 
consecutive seasons (2019, 2020 and 2021).  After recalling the main parameters of TWA~7 in Sec.~\ref{sec:par}, we detail the recorded observations in Sec.~\ref{sec:obs}, 
investigate the rotational modulation of magnetic and temperature proxies in Sec.~\ref{sec:mag}, study the large-scale and small-scale magnetic topologies at the surface of 
TWA~7 with Zeeman-Doppler imaging (ZDI) in Sec.~\ref{sec:zdi}, and outline the RV and activity measurements with their modeling in Secs.~\ref{sec:rvs} and \ref{sec:act}.  
We finally summarize and discuss the implication of our results for our understanding of star and planet formation in Sec.~\ref{sec:dis}.

\section{The young M dwarf TWA~7} 
\label{sec:par}

TWA~7 (CE~Ant) is a PMS M2 dwarf located at $34.01\pm0.03$~pc \citep{Gaia20} in TWA and aged $10\pm2$~Myr \citep{Luhman23}.  
Classified as a weak-line, non-accreting T-Tauri star, TWA~7 has a photospheric temperature of $\teff=3509\pm116$~K, a radius of $\rstar=0.92\pm0.12$~\rsun, 
and a mass of $\mstar=0.46\pm0.09$~\msun, yielding a surface gravity $\logg=4.18\pm0.18$~dex (with $g$ in \cmss) and a luminosity of $\lstar=0.115\pm0.019$~\lsun\ \citep{Stassun18}.  
In this reference paper, data from different surveys were assembled and matched with PHOENIX stellar evolution models \citep{Husser13} to derive reliable stellar parameters in 
a homogeneous way for a vast ensemble of stars.  Consistent stellar parameters were derived from our SPIRou nIR spectra (see Sec.~\ref{sec:obs}).  
Matching the quoted temperature and radius with other evolution models yields slightly different masses, $0.40\pm0.08$~\msun\ for 
models of \citet{Baraffe15} and non-magnetic models of \citet{Feiden16}, and $0.54\pm0.12$~\msun\ for magnetic models of \citet{Feiden16}, both in agreement within error bars 
with the previous estimate.  In this picture, TWA~7 is fully convective and still contracting towards the main-sequence, appearing as a younger and slightly lower mass version of 
the nearby active, planet-hosting star AU~Mic.  {\emr Similar atmospheric properties were inferred} in several independent studies \citep[e.g.,][]{Mentuch08,Lopez-Valdivia23}, though 
not all of them agree \citep[e.g.,][]{Yang08,Nicholson21,Paolino25}.  

Photometric observations revealed that TWA~7 has a rotation period of $5.00\pm0.03$~d \citep{Lawson05,Messina10}, with a modulation of amplitude $\simeq$10~mmag.  More recently, 
TWA~7 was monitored with TESS in 2019, 2021, 2023, and 2025, with data in the first three seasons showing modulation at the rotation period and additional stochastic fluctuations 
due to stellar flares or to satellite issues (in 2025).  As a result of its relatively rapid rotation, TWA~7 was found to be strongly magnetic, with a small-scale field ranging 
from 2.2 to 2.7~kG \citep{Yang08,Lavail19,Lopez-Valdivia23}.  

\begin{table}[t!]
\caption{Parameters of TWA~7 used in our study} 
\centering
\begin{tabular}{ccc}
\hline
distance (pc)    & $34.01\pm0.03$  & \citet{Gaia20}    \\ 
\teff\ (K)       & $3435\pm60$     & this paper \\ 
\logg\ (\cmss)   & $4.28\pm0.10$   & this paper \\ 
$\rm [Fe/H]$     & $0.21\pm0.10$   & this paper        \\ 
\mstar\ (\msun)  & $0.46\pm0.09$   & \citet{Stassun18} \\ 
\rstar\ (\rsun)  & $0.92\pm0.12$   & \citet{Stassun18} \\ 
\Prot\ (d)       & $5.012\pm0.007$ & this paper        \\ 
$i$ (\degr)      & 13              & disk incl., used for ZDI   \\ 
\vsini\ (\kms)   & $2.1\pm0.3$     & from \Prot, \rstar, and $i$  \\ 
<$B$> (kG)       & $3.2\pm0.3$     & this paper        \\ 
\hline
\end{tabular}
\label{tab:par}
\end{table} 

TWA~7 is surrounded by a complex debris disk viewed almost face-on, inclined at only $i\simeq$13\degr\ to the line of sight \citep{Ren21}.  The debris disk of TWA~7 features 
several substructures, including distinct disk components, spiral arms, and a southern dust clump \citep{Ren21}. The middle disk component, a very narrow ring located 
between the broad inner and outer disk rings, had been interpreted as a possible resonant structure triggered by an unseen planet companion.  Recent observations revealed 
the presence of a sub-jovian candidate planet at the predicted location in the debris disk \citep[about 52~au from the host star;][]{Lagrange25,Crotts25}.  

Assuming that the stellar equatorial plane coincides with that of the disk implies that the spectral lines of TWA~7 are only weakly broadened by rotation despite the short 
\Prot, with a line-of-sight projected equatorial rotation velocity $\vsini=2.1\pm0.3$~\kms.  The main stellar parameters used in our study are summarized in Table~\ref{tab:par}.

\section{SPIRou observations}
\label{sec:obs}

We observed TWA~7 between 2019 February~14 and 2021 May~02, with the SPIRou nIR spectropolarimeter \citep{Donati20} at CFHT, within 
the SPIRou Legacy Survey (RUNIDs 19AP40, 20AP40, and 21AP40, PI J.-F.~Donati).  SPIRou collects unpolarized and polarized stellar spectra, covering a 
wavelength interval of 0.95--2.50~\mic\ at a resolving power of 70\,000 in a single exposure.  Each observation usually consists of a sequence of four sub-exposures, with 
each sub-exposure corresponding to a different azimuth of the Fresnel rhomb retarders of the SPIRou polarimetric unit.  With this procedure, we are able to remove systematics 
in polarization spectra to first order \citep[][]{Donati97b}.  Each recorded sequence yields one unpolarized (Stokes $I$) and one circularly polarized (Stokes $V$) 
spectrum, as well as one null polarization check (called $N$) used to diagnose potential instrumental or data reduction issues.  We collected a total of 52 polarization sequences 
of TWA~7 over three consecutive observing seasons.  One of them, obtained on 2020 Jan~26, was affected by an instrumental issue and discarded from the set, leaving us with a 
time series of 51 spectra spread over 808~d, 17 in 2019, 8 in 2020 and 26 in 2021.  Total exposure times 
ranged from 670~s (in early 2019) to 1200~s in 2020 and 2021, while signal to noise ratios (S/Ns) per 2.3~\kms\ pixel in the $H$ band spread from 134 to 365 (median 280).  

\begin{figure}[ht!]
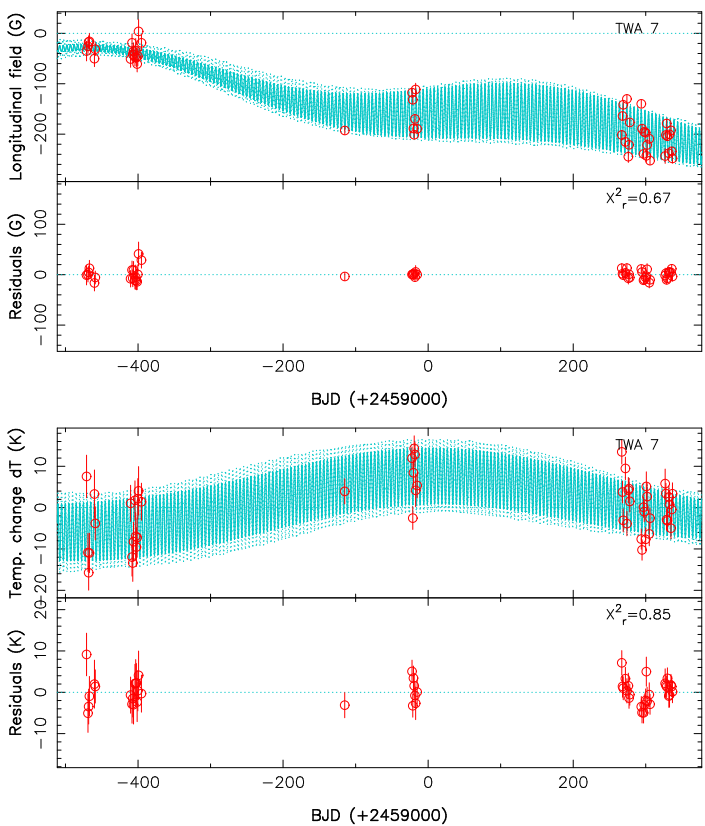

\centerline{\includegraphics[scale=0.39,angle=-90]{fig/twa7-gpb.ps}\vspace{2mm}}
\centerline{\includegraphics[scale=0.39,angle=-90]{fig/twa7-gpt.ps}}
\caption[]{Longitudinal magnetic field \Bl\ (top panel) and temperature variations $dT$ (bottom panel) of TWA~7 (red dots), and QP GPR fit to the data (cyan full line) with corresponding 
68~percent confidence intervals (cyan dotted lines).  The residuals, shown in the bottom plot of each panel, yield rms of 8.2~G and 2.9~K ($\chisqr=0.67$ and 0.85).  
{\emr (March 1st 2019, 2020 and 2021 at noon correspond to x-values of $-$456, $-$90 and 275, respectively).}  A zoom on the 2021 data is shown in Fig.~\ref{fig:gpb2}.  } 
\label{fig:gpb}
\end{figure}

We processed all spectra with \texttt{Libre ESpRIT}, the nominal reduction pipeline of ESPaDOnS at CFHT, optimized for spectropolarimetry and adapted for SPIRou \citep{Donati20}.  
We then applied least-squares deconvolution \citep[LSD;][]{Donati97b} to the reduced spectra, using a line mask computed with the VALD-3 database \citep{Ryabchikova15} for a set 
of atmospheric parameters close to those of TWA~7 ($\teff=3500$~K, $\logg=4.5$~\cmss).  We only selected atomic lines deeper than 10\% of the continuum level, for a total of 
$\simeq$1500 lines of average wavelength and Land\'e factor equal to 1750~nm and 1.2.  This yielded noise levels, $\sigma_V$, in the resulting Stokes $V$ LSD profiles ranging 
from 2.3 to 7.5 (median 3.1, in units of $10^{-4} I_c$ where $I_c$ denotes the continuum intensity).  Zeeman signatures were well detected, especially in 2020 and 2021, with 
average peak-to-peak amplitudes of 0.13\%, 0.52\%, and 0.76\% in 2019, 2020 and 2021.  We computed the longitudinal field, \Bl, defined as the line-of-sight-projected vector 
magnetic field at the stellar surface averaged over the visible hemisphere, from LSD Stokes $IV$ profiles following \citet{Donati97b}, integrating over an interval of $\pm$30~\kms\ 
adequate for TWA~7.  We found that \Bl\ is unambiguously detected, with a reduced chi square (relative to \Bl=0) of \chisqr=332.  The same operation applied to the null polarization 
spectrum $N$ yielded \chisqr=0.96, consistent with no spurious signal down to the noise level and indicating no issues in the observation and reduction procedures.  \Bl\ was mostly 
negative in TWA~7, ranging from $-$253 to 4~G with a median error bar of 9.4~G (see Fig.~\ref{fig:gpb}), and exhibited moderate yet clear rotational modulation (see 
Sec.~\ref{sec:mag}).  

We also reduced our SPIRou data with the latest version of \texttt{APERO} (v0.7.294), the nominal SPIRou reduction pipeline \citep{Cook22} optimized for RV precision.  We analyzed the  
\texttt{APERO} spectra with the line-by-line (LBL) technique \citep[v0.65;][]{Artigau22}, yielding accurate RVs and differential temperatures $dT$ estimated from the variation of spectral 
lines with respect to their median profile \citep{Artigau24}.   We corrected these RVs for spectrograph drifts using the Fabry-Perot spectrum that SPIRou records simultaneously with the 
stellar spectrum \citep{Donati20}.  We found that RVs ranged from $-$17 to 25~\ms\ with a median error bar of 1.4~\ms, while $dT$ varied from $-$16 to 14~K with 
a median error bar of 3.0~K.  We then used the $dT$ measurements to infer a relative photometric light curve at SPIRou wavelengths (with temperature changes converted into brightness 
fluctuations with the Planck function), to be adjusted with ZDI along with the LSD profiles (see Sec.~\ref{sec:zdi}).  We obtained light curves at SPIRou wavelengths 
with full amplitudes of $\simeq$1\%, similar to, and in phase with, those measured with TESS in the $I$ band at similar epochs (2019 March and 2021 March).  {\emr In addition to 
constraining the spot pattern, $dT$ also indirectly informs on the small-scale field at the surface of TWA~7 and its variation with time via the tight correlation reported between both 
quantities for M dwarfs \citep{Artigau24,Cristofari25}.}.  

The full log of our observations is provided in Table~\ref{tab:log}.  Phases and rotation cycles are derived assuming a rotation period of $\Prot=5.012$~d (see Table~\ref{tab:par}), 
counting from an arbitrary starting barycentric Julian date (BJD) of 2\,458\,529 (slightly before our first observation of TWA~7).  

Finally, we used our median unpolarized spectrum of TWA~7 to double check the atmospheric parameters with ZeeTurbo, {\emr a spectrum modeling tool specifically designed for 
characterizing nIR spectra of M dwarfs by comparing them to synthetic spectra computed from MARCS model atmospheres, including the Zeeman broadening of spectral lines induced by 
small-scale magnetic fields \citep[][]{Cristofari23}. }  For the stellar parameters, we found $\teff=3435\pm60$~K, $\logg=4.28\pm0.10$, and a metallicity relative to the Sun, 
[Fe/H]~=~$0.21\pm0.10$, in good agreement with \citet{Stassun18}.  We also obtained that TWA~7 hosts a small-scale magnetic field, <$B$>~=~$3.2\pm0.3$~kG, {\emr with the median stellar 
spectrum described as a linear combination of synthetic spectra with fields of strengths 0, 2, 4, 6, 8, and 10~kG components respectively covering} relative fractions of the visible 
hemisphere equal to 0.18, 0.44, 0.17, 0.08, 0.05, and 0.08 (with typical error bars of a few 0.01).  In this analysis, we fixed $\vsini=2.1$~\kms\ (see Table~\ref{tab:par}) while 
radial-tangential macroturbulence $\zeta$ was adjusted with all other parameters, yielding $\zeta=4.6\pm0.1$~\kms.  We show in Fig.~\ref{fig:spc} the achieved fit to the K band of the spectrum.  

Previous studies derived a weaker small-scale field, of 2.3~kG \citep[][\emr data from 2000]{Yang08} and 2.7~kG \citep[][\emr data from 2008 and 2009]{Lavail19}, with a similar technique 
limited to fewer model components (0, 2, 4, and 6~kG for 
\citealt{Yang08}, and 0, 2, and 4~kG for \citealt{Lavail19}), stronger ones being unnecessary on the grounds of the Bayesian information content \citep[BIC,][]{Lavail19}.  Our observations 
however suggest that the strongest field components are required, yielding a smaller BIC value when included.  This possibly suggests that the small-scale field changed between the different 
measurement epochs.  Alternatively, the far wings of the magnetically sensitive lines of our median TWA~7 spectrum may suffer normalization issues, rendering the additional small-scale field 
from the highest field components less reliable.  We did not detect rotational modulation of <$B$> by looking at individual spectra of TWA~7.  This is not surprising given the low value of $i$ 
and the low amplitude modulation of $dT$ (see Sec.~\ref{sec:mag}), suggesting a modulation amplitude of order 0.1~kG for <$B$> (judging from the correlation between $dT$ and <$B$>).  

\begin{table}[t!]
\caption{Results of the MCMC modeling of the \Bl\ (top) and $dT$ (bottom) curves of TWA~7}
\centering
\resizebox{\linewidth}{!}{
\begin{tabular}{cccc}
\hline
Parameter   & Symbol & Value & Prior   \\
\hline
\Bl &&&  \\
GP amplitude (G)     & $\theta_1$  & $94\pm20$        & mod Jeffreys ($\sigma_{\Bl}$) \\
Rec.\ period (d)     & $\theta_2$  & $5.012\pm0.007$  & Gaussian (5.0, 1.0) \\
Evol.\ timescale (d) & $\theta_3$  & $240\pm54$       & log Gaussian ($\log$ 250, $\log$ 2) \\
Smoothing            & $\theta_4$  & 2.0              & fixed \\
White noise (G)      & $\theta_5$  & $3.7\pm2.4$      & mod Jeffreys ($\sigma_{\Bl}$) \\
Rms (G)              &             & 8.2              & \\
$\chisqr$            &             & 0.67             & \\
\hline
$dT$ &&&  \\
GP amplitude (K)     & $\theta_1$  & $10.6\pm2.5$     & mod Jeffreys ($\sigma_{dT}$) \\
Rec.\ period (d)     & $\theta_2$  & $5.016\pm0.010$  & Gaussian (5.0, 1.0) \\
Evol.\ timescale (d) & $\theta_3$  & $257\pm91$       & log Gaussian ($\log$ 250, $\log$ 2) \\
Smoothing            & $\theta_4$  & 1.3              & fixed  \\
White noise (K)      & $\theta_5$  & $1.5\pm0.8$      & mod Jeffreys ($\sigma_{dT}$) \\
Rms (K)              &             & 2.9              & \\
$\chisqr$            &             & 0.85             & \\
\hline
\end{tabular}}
\tablefoot{For each hyper parameter, we list the fitted value along with the corresponding error bar and the assumed prior.  The knee of the modified Jeffreys prior is set to
the median error bars of \Bl\ and $dT$ (9.4~G and 3.0~K).  For the evolution timescale,  $\theta_3$, the log Gaussian prior is set to 250~d (within a factor of 2), 
following a preliminary run. The smoothing parameter, $\theta_4$, weakly constrained by the \Bl\ and $dT$ data, was fixed to the value that maximized likelihood in a previous MCMC run.}
\label{tab:gpr}
\end{table}

\section{Magnetic field \& temperature changes}
\label{sec:mag}

We used the framework of \citet{Haywood14} and \citet{Rajpaul15} to perform a quasiperiodic (QP) Gaussian process regression (GPR) fit to the \Bl\ values, arranged in a vector denoted $y$. 
In this purpose, we employed the following QP covariance function $c(t,t')$:
\begin{eqnarray}
c(t,t') = \theta_1^2 \exp \left( -\frac{(t-t')^2}{2 \theta_3^2} -\frac{\sin^2 \left( \frac{\pi (t-t')}{\theta_2} \right)}{2 \theta_4^2} \right) 
\label{eq:covar}
\end{eqnarray}
where $\theta_1$ is the amplitude (in G) of the Gaussian process (GP), $\theta_2$ its recurrence period (measuring \Prot),
$\theta_3$ the evolution timescale on which the \Bl\ curve changes shape (in d), and $\theta_4$ is a smoothing parameter describing the amount of allowed harmonic complexity 
{\emr (the smaller $\theta_4$, the larger the complexity, with $\theta_4>1$ implying a nearly sinusoidal fluctuation over a timescale of $\theta_3$)}.
A fifth hyperparameter, $\theta_5$, describes the excess uncorrelated noise required to obtain the QP GPR fit to the \Bl\ data with the highest
likelihood $\mathcal{L}$, defined by
\begin{eqnarray}
2 \log \mathcal{L} = -n \log(2\pi) - \log|C+\Sigma+S| - y^T (C+\Sigma+S)^{-1} y
\label{eq:llik}
\end{eqnarray}
where $C$ is the covariance matrix for all observing epochs, $\Sigma$ is the diagonal variance matrix associated with $y$, $S=\theta_5^2 J$ is the contribution of the additional
white noise, with $J$ the identity matrix, and $n$ the number of data points.

With a Monte-Carlo Markov Chain (MCMC) process, we explored the hyperparameter domain, yielding posterior distributions and error bars for
each.  The MCMC and GPR modeling tools are those used in our previous studies \citep[e.g.,][]{Donati23}.  The MCMC process is based on a conventional single chain Metropolis-Hastings scheme, 
typically running over a few $10^5$ steps, including the first few $10^4$ steps as burn-in.  Convergence was checked with an autocorrelation analysis, verifying that the burn-in and main 
phase vastly exceed the autocorrelation lengths of all parameters.  As in \citet[][]{Haywood14}, we computed the marginal logarithmic likelihood $\log \mathcal{L}_M$ of a given solution 
following \citet{Chib01}.

The GPR fit we obtained is shown in the top panel of Fig.~\ref{fig:gpb}, with a zoom on the 2021 data in Fig.~\ref{fig:gpb2}.  The derived GPR hyperparameters are listed in the top 
section of Table~\ref{tab:gpr}.  We found that \Bl\ variations came mostly from the seasonal evolution of the large-scale field between 2019 and 2020.  Rotational modulation was moderate, as 
expected from the low inclination of the stellar rotation axis to the line of sight, but is nonetheless clearly detected in 2020 and 2021, with a typical semi-amplitude of $\simeq$50~G.  

We performed a similar analysis on the $dT$ measurements (see bottom panel of Fig.~\ref{fig:gpb} and bottom section of Table~\ref{tab:gpr}).  
Rotational modulation dominated the temperature variations of TWA~7 during our observations, and is clearly detected with a typical semi-amplitude 
of 5-7~K.  The seasonal changes were weaker, suggesting that the spot pattern and the underlying small-scale field <$B$> did not vary much over our campaign.  This contrasts with the 
large seasonal \Bl\ variations over the same period, suggesting a year-to-year change in the large-scale field topology but not in its small-scale intensity (\Bl\ being very sensitive 
to field orientation as opposed to <$B$>).  As for AU~Mic \citep[][]{Donati25}, $dT$ and \Bl\ are poorly correlated (Pearson's coefficient $\simeq$$-0.3$).  

The recurrence period derived for \Bl\ and $dT$ agree well, and match literature values \citep{Lawson05,Messina10} with smaller error bars.

\section{ZDI modeling}
\label{sec:zdi}

We used our LSD Stokes $IV$ profiles in each of the three seasons, in addition to the photometric constraints derived from $dT$, to derive maps of the large-scale magnetic field and photospheric 
brightness at the surface of TWA~7 with ZDI.  Including both LSD Stokes $I$ and $V$ (rather than Stokes $V$ only) profiles in the ZDI modeling is essential to ensure that the Zeeman broadening 
of line profiles is well reproduced and that the reconstructed field is not severely underestimated \citep[e.g.,][]{Donati25b}.  In practice, ZDI operates as outlined in previous studies.  
Beginning with empty magnetic maps and brightness distributions, ZDI iteratively adds information on both images, exploring the parameter space with conjugate gradient techniques and comparing the 
synthetic Stokes profiles of the current images with observed ones at each iteration, until it reaches the requested level of agreement with the data (i.e., a given \chisqr).

We divided the surface of TWA~7 into 5000 grid cells, associating each with a value of the local surface brightness (relative to the quiet photosphere).  We described the magnetic 
topology with a spherical harmonics (SH) expansion using the formalism of \citet{Donati06b} in its revised implementation \citep{Lehmann22, Finociety22}, where the poloidal and toroidal 
components of the vector field are expressed with three sets of complex SH coefficients, $\alpha_{\ell,m}$ and $\beta_{\ell,m}$ for the poloidal component, and $\gamma_{\ell,m}$ for the toroidal 
component (with $\ell$ and $m$ denoting the degree and order of the corresponding SH term in the expansion).  Given the low \vsini\ of TWA~7, we limited the SH expansion to $\ell=5$.  With more 
parameters than independent data points, the inversion problem is ill-posed, requiring some regularization.  ZDI chooses the simplest of all solutions, i.e., the one with minimum 
information or maximum entropy that matches the data at the requested \chisqr\ level, following the approach of \citet[][]{Skilling84}.  {\emr We assumed TWA~7 rotates as a solid body.  
Although latitudinal differential rotation is in principle accessible to ZDI \citep[e.g.,][]{Finociety23}, measuring it on stars viewed nearly pole-on with low \vsini\ like TWA~7 is hardly 
feasible and not reliable.} 

\begin{figure*}[ht!]
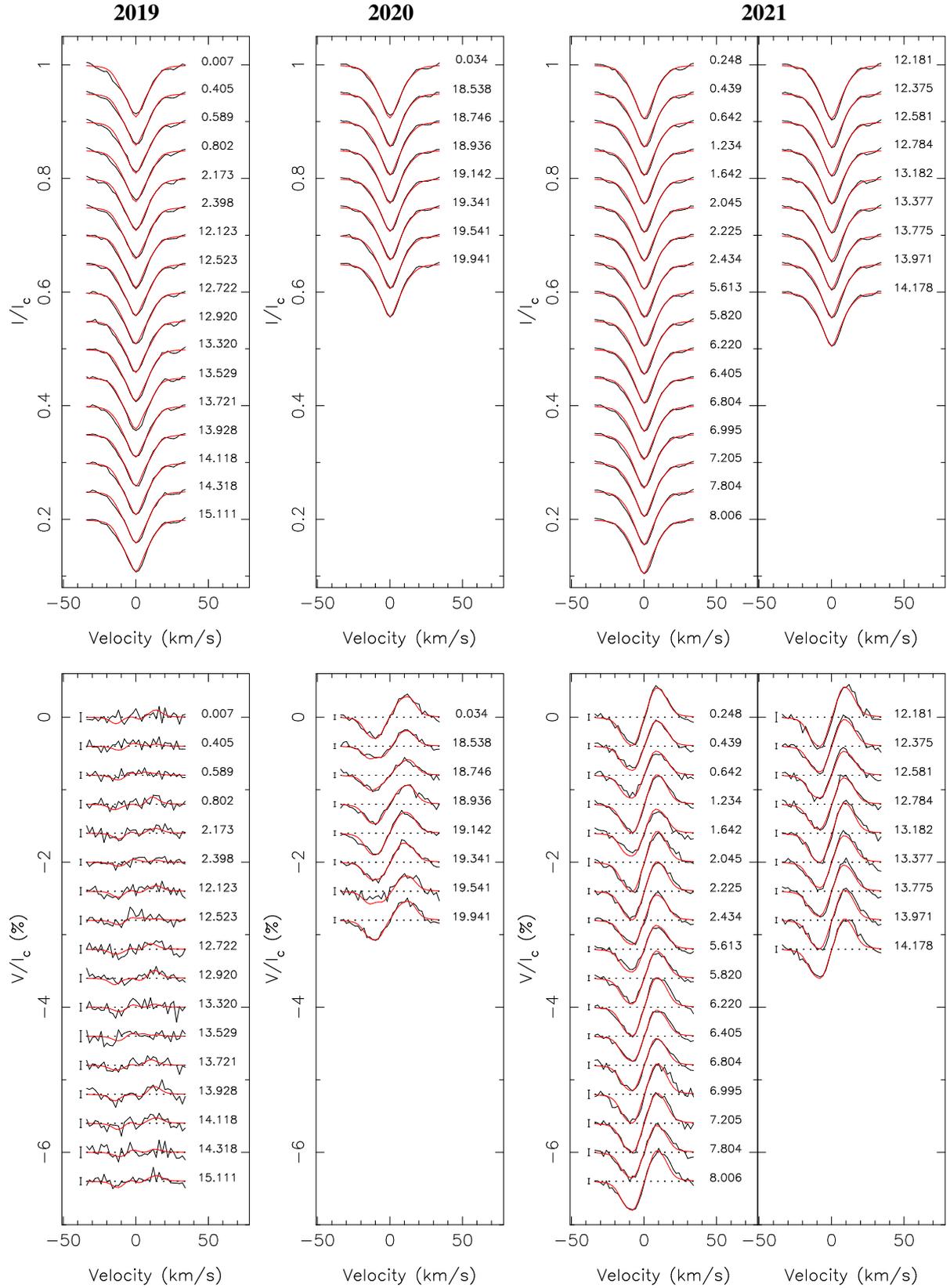
 
\flushleft{\large\bf \hspace{3.1cm}2019\hspace{3.5cm}2020\hspace{5.5cm}2021\vspace{-4mm}}
\center{\includegraphics[scale=0.31,angle=-90]{fig/twa7-fiti19.ps}\hspace{2mm}\includegraphics[scale=0.31,angle=-90]{fig/twa7-fiti20.ps}\hspace{2mm}\includegraphics[scale=0.31,angle=-90]{fig/twa7-fiti21.ps}\vspace{1mm}}   
\center{\includegraphics[scale=0.31,angle=-90]{fig/twa7-fitv19.ps}\hspace{2mm}\includegraphics[scale=0.31,angle=-90]{fig/twa7-fitv20.ps}\hspace{2mm}\includegraphics[scale=0.31,angle=-90]{fig/twa7-fitv21.ps}}   
\caption[]{Observed (thick black line) and modelled (thin red line) LSD Stokes $I$ (top row) and $V$ (bottom row) profiles of TWA~7, for seasons 2019, 2020 and 2021 (from left to right).
Rotation cycles (counting from 0, 71, and 147, for seasons 2019, 2020, and 2021, respectively, see Table~\ref{tab:log}) are indicated to the right of LSD profiles, and $\pm$1$\sigma$ error bars to the left 
of Stokes $V$ profiles. }
\label{fig:fit}
\end{figure*}

\begin{figure*}[ht!]
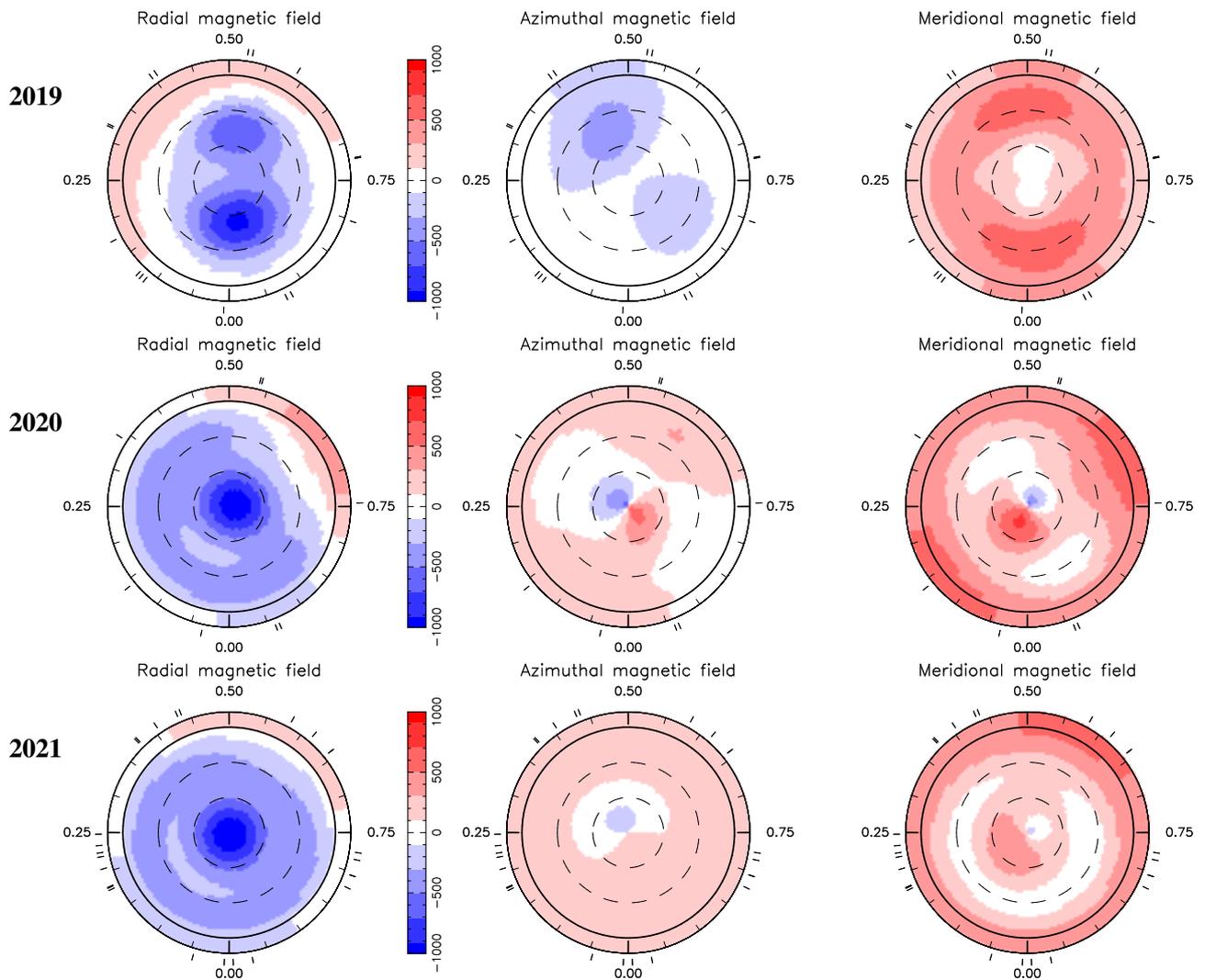
 
\centerline{\large\bf 2019\raisebox{0.3\totalheight}{\includegraphics[scale=0.45,angle=-90]{fig/twa7-map19.ps}}\vspace{1mm}}
\centerline{\large\bf 2020\raisebox{0.3\totalheight}{\includegraphics[scale=0.45,angle=-90]{fig/twa7-map20.ps}}\vspace{1mm}}
\centerline{\large\bf 2021\raisebox{0.3\totalheight}{\includegraphics[scale=0.45,angle=-90]{fig/twa7-map21.ps}}} 
\caption[]{Reconstructed maps of the large-scale field of TWA~7 showing the radial, azimuthal and meridional components in spherical 
coordinates (left, middle and right columns, units in G), for season 2019, 2020 and 2021 (top to bottom rows, respectively).  
These maps, derived from the LSD Stokes $IV$ profiles of Fig.~\ref{fig:fit} using ZDI, are displayed in a flattened polar projection 
down to latitude $-13$\degr, with the north pole at the center and the equator depicted as a bold line.  Outer ticks mark the phases of
observations.  Positive radial, azimuthal, and meridional fields point outwards, counterclockwise, and polewards, respectively. }
\label{fig:map}
\end{figure*} 

\begin{table}[t!]
\caption{Properties of the large-scale and small-scale field {\emr for the magnetic topologies reconstructed with ZDI}, for all three seasons}
\centering
\resizebox{\linewidth}{!}{
\begin{tabular}{ccccccc}
\hline
         & \multicolumn{6}{c}{Stokes $IV$ analysis}                      \\
         & \multicolumn{6}{c}{($f_I=0.8$, $f_V=0.2$, $\vD=2.0$~\kms)}    \\
\hline
Season   & <$B_V$> & <$B_I$> & <$B_s$> & \Bd  & tilt / phase & pol/axi   \\
         &  (kG)   & (kG)    & (kG)   & (kG) & (\degr / ) & (\%) \\
\hline
2019     & 0.45   & 1.8   & 2.2   & $-$0.48 & 22 / 0.88 & 95 / 87 \\
2020     & 0.51   & 2.0   & 2.2   & $-$0.62 & 19 / 0.13 & 92 / 89 \\
2021     & 0.52   & 2.1   & 2.1   & $-$0.69 & 15 / 0.08 & 92 / 95 \\
\hline
\end{tabular}}
\tablefoot{ Columns 2 and 3 respectively list the large-scale and small-scale fields, once quadratically averaged over the {\emr reconstructed maps}.  
Column~4 gives the time-averaged small-scale field integrated over the visible hemisphere <$B_s$>.  
Columns~5 to 7 respectively list the polar strengths of the dipole component \Bd, the tilt of the dipole component to the rotation axis and the phase towards 
which it is tilted, and the amount of magnetic energy reconstructed in the poloidal component of the field and in the axisymmetric modes of this component.  
Error bars on field values and percentages are equal to $\simeq$10\%, whereas dipole tilts are accurate to $\simeq$5\degr. }
\label{tab:mag}
\end{table}

We used Unno-Rachkovsky's equation of the polarized radiative transfer equation in a plane-parallel Milne-Eddington atmosphere \citep{Landi04} to compute local synthetic Stokes $IV$ profiles 
for each grid cell.  At each observed rotation phase, we summed the spectral contributions from all visible cells (assuming a linear center-to-limb darkening law for the continuum, with a 
coefficient of 0.3) to obtain the overall synthetic profiles, whose mean wavelength and Land\'e factor are mirrored from those of our LSD profiles, i.e., 1750~nm and 1.2.  We assumed a 
Doppler width of the local profile of $\vD=2.0$~\kms, yielding the best fit to the Stokes $I$ profiles.  As in previous studies, we introduced a filling factor for the large-scale field, 
$f_V$, and another one for the small-scale field, $f_I$, both assumed constant over the star.  It implies that each cell with a reconstructed field $B$ hosts a field strength $B/f_V$ in a 
fraction of the cell equal to $f_V$ and $f_I$ for Stokes $V$ and $I$ profiles, respectively.  For this study, we set $f_I=0.8$ (consistent with the results of ZeeTurbo, see Sec.~\ref{sec:obs}) 
and $f_V=0.2$ \citep[as for other similar young stars, including AU~Mic;][]{Donati25}.  We computed synthetic light curves in a similar way, by summing the photometric contributions of 
all visible cells, estimated from their local brightness and limb angle at each observed rotation phase.  

\begin{table*}[t!]
\caption{MCMC results of the RV modeling}
\centering
\begin{tabular}{cccccc}
\hline
Parameter          & No planet                 & Planet b        &  Planet b'           &  Planet b"          & Prior \\
                   &                           & (15.2~d)        &  (20.8~d)            &  (30.4~d)           &       \\
\hline
$\theta_1$ (\ms)   & $8.0^{+3.7}_{-2.5}$  & $10.4^{+3.0}_{-2.3}$ & $9.7^{+2.9}_{-2.2}$  & $9.9^{+2.7}_{-2.1}$  & modified Jeffreys ($\sigma_{\rm RV}$) \\
$\theta_2$ (d)     & $5.02\pm0.15$        & $4.96\pm0.11$        & $5.05\pm0.06$        & $5.03\pm0.09$        & Gaussian (5.0; 0.3) \\
$\theta_3$ (d)     & $24^{+11}_{-8}$      & $28^{+9}_{-7}$       & $48^{+18}_{-13}$     & $29^{+9}_{-7}$       & log Gaussian ($\log$ 40; $\log$ 2.0) \\
$\theta_4$         & 1.3                  & 1.3                  & 1.3                  & 1.3                  & fixed \\ 
$\theta_5$ (\ms)   & $6.1\pm1.1$          & $4.1\pm0.8$          & $5.1\pm0.8$          & $4.5\pm0.8$          & modified Jeffreys ($\sigma_{\rm RV}$) \\
\hline
$K_b$ (\ms)        &                      & $5.4^{+1.3}_{-1.0}$  & $4.8^{+1.5}_{-1.2}$  & $7.0^{+2.3}_{-1.7}$  & modified Jeffreys ($\sigma_{\rm RV}$) \\
$P_b$ (d)          &                      & $15.21\pm0.02$       & $20.83\pm0.05$       & $30.35\pm0.10$       & Gaussian (15.2, 20.8 or 30.3; 0.3)        \\ 
$T_b$ (2459000+)   &                      & $289.1\pm0.6$        & $295.1\pm1.0$        & $276.3\pm1.2$        & Gaussian (289, 295, or 276; 5)            \\ 
$M_b \sin i$ (\me) &                      & $12.5^{+3.0}_{-2.3}$ & $12.3^{+3.8}_{-3.1}$ & $20.4^{+6.7}_{-4.9}$ & from $K_b$, $P_b$, \mstar \\
$M_b$ (\mjup)      &                      & $0.17^{+0.04}_{-0.03}$ & $0.17^{+0.05}_{-0.04}$ & $0.28^{+0.09}_{-0.07}$ & assuming $i=13$\degr  \\
$a_b$ (au)         &                      & $0.092\pm0.006$      & $0.114\pm0.007$      & $0.147\pm0.009$      & \\ 
\hline
\chisqr            & 10.6                 & 3.8                  & 7.1                  & 5.1                  &   \\
rms (\ms)          & 4.7                  & 2.8                  & 3.8                  & 3.3                  &   \\
$\log \mathcal{L}_M$ & 174.9              & 186.3                & 182.9                & 183.6                &   \\
$\Delta \log \mathcal{L}_M$ & 0.0         & 11.4                 & 8.0                  & 8.7                  &   \\
\hline
\end{tabular}
\tablefoot{We list the recovered GP and {\emr parameters of the tentative planet} with their error bars, as well as the priors used where relevant, 
for the model without planet (column 2) and the best models with a planet (column 3 to 5).  The last four rows report the \chisqr, the rms of the best fit to the RV data, 
the associated marginal logarithmic likelihood, $\log \mathcal{L}_M$, and the marginal logarithmic likelihood variation, $\Delta \log \mathcal{L}_M$, relative to the case without planet.}
\label{tab:pla}
\end{table*}

The fits to the LSD Stokes $IV$ profiles of TWA~7 derived with ZDI for each of the three epochs are shown in Fig.~\ref{fig:fit}, the corresponding reconstructed maps are presented in 
Fig.~\ref{fig:map}, and their main characteristics listed in Table~\ref{tab:mag}.  The brightness maps we simultaneously recovered only showed low-contrast features, in agreement with the 
low-amplitude photometric curves derived from $dT$ measurements (see Fig.~\ref{fig:pho}).  

The derived magnetic topology was mostly poloidal at all epochs, with the dipole component encompassing 82$-$89\% of the reconstructed poloidal field energy, and resembled that inferred for 
AU~Mic \citep{Donati25,Donati25b}.  The main topological change occurred between 2019 and 2020, as expected from the large increase in the amplitude of LSD Stokes $V$ signatures (see 
Fig.~\ref{fig:fit}).  Throughout the observing campaign, the reconstructed field of TWA~7 got simpler, less tilted with respect to the rotation axis, and more dipolar.  The polar strength 
of the dipole increased from 0.48 to 0.69~kG from 2019 to 2021, while its polarity remained negative.  The toroidal field was weak (5$-$8\% of the reconstructed field energy), close to the 
detection level, and switched polarity between 2019 and 2021 a potential signature of a cyclic evolution of the large-scale field.  

The small-scale field <$B_s$> inferred with ZDI (2.2~kG) remained more or less stable throughout the campaign, as did $dT$ (see Sec.~\ref{sec:mag}), presumably a good proxy for <$B$> 
\citep{Artigau24,Cristofari25}.  The ZDI analysis thus confirmed the preliminary conclusions of Sec.~\ref{sec:mag} that the magnetic field evolution of TWA~7 over our observing period 
mostly concerned the large-scale topology rather than the average strength of the small-scale field.  
We note that <$B_s$> is weaker than the <$B$> estimate directly derived with ZeeTurbo ($3.2\pm0.3$~kG) for some unclear reason, despite numerous attempts to reach a better 
agreement by tuning the model parameters.  It suggests that ZDI did not fully succeed at fitting the farthest wings of the LSD profiles of TWA~7.  We had no such issue in our previous study 
on magnetic topologies of two active M dwarfs from all Stokes parameters \citep{Donati25b}, where small-scale fields were either similar in strength or even stronger than that of TWA~7, 
but with a different distribution of field strengths and filling factors over the surface.  The derived <$B_s$> is nonetheless consistent with some of the small-scale field measurements 
previously reported for TWA~7 \citep{Yang08}.

\section{RV modeling}
\label{sec:rvs}

We analyzed the RVs of TWA~7 inferred by LBL, looking for a potential RV signal from a close-in giant planet.  We proceeded as in Sec.~\ref{sec:mag}, using GPR to model the activity jitter 
in the RV curve and a simple sine wave to describe the RV signature of the putative close-in planet (assumed to be on a circular orbit).  This implied a total of seven parameters to be 
fitted, four for the activity jitter (with $\theta_4$ fixed as for $dT$) and three for the close-in planet (the orbital period $P_b$, the semi-amplitude of the RV signature $K_b$, and the 
BJD of the second conjunction $T_b$).  Posterior distributions for all parameters are derived with a MCMC process, while the robustness of a given planet signal is estimated from the 
variation in marginal likelihood $\Delta \log \mathcal{L}_M$ relative to the model with no planet.  We did not take into account the newly discovered distant planet in our modeling, its 
orbital period (of about 550~yr for a semi-major axis of 52~au) being much longer than that of our observations.  

\begin{figure}[ht!]
\centerline{\includegraphics[scale=0.39,angle=-90]{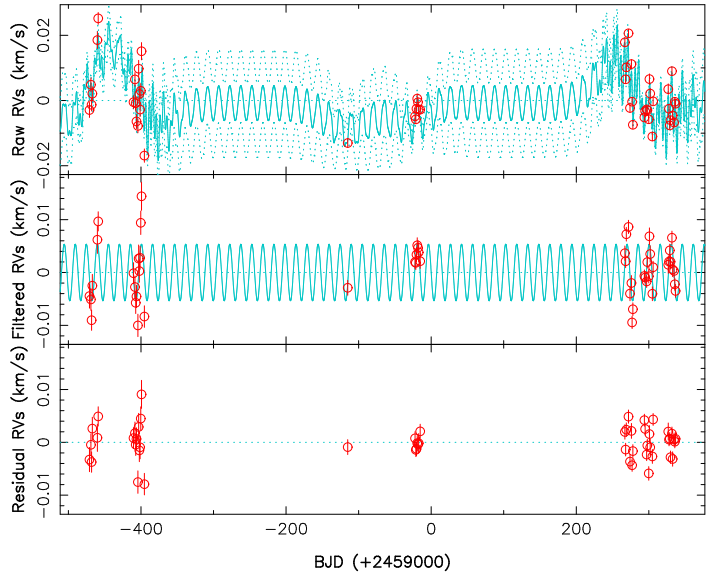}}
\caption[]{Raw (top), filtered (middle), and residual (bottom) RVs of TWA~7 (red dots) over the observing period.  The top panel shows the MCMC fit to the data, including a QP GPR modeling 
of the activity and a {\emr tentative planet} on a 15.21-d circular orbit (cyan).  The middle panel shows the {\emr tentative planet} RV signature (cyan) once activity is filtered out.  
The rms of the residuals is 2.8~\ms.  A zoom on the 2021 data is shown in Fig.~\ref{fig:rv2}.}
\label{fig:rv}
\end{figure}

We started by modeling activity, showing up clearly in the RV data at a level of 5$-$10~\ms, and noticed that, once removed, some power was still present in the periodogram of RV residuals 
around periods of 15, 20 and 30~d (and their 1-yr and 2-yr aliases).  
We limited our analysis to periods up to $\simeq$50~d, i.e., those showing the strongest peaks and the only ones we could reliably investigate given the moderate sampling 
of our data in individual seasons.  We first used a wide prior for $P_b$ to find out which period among all aliases was the most likely, then narrow priors to focus on the main peaks.  We 
found that the strongest signal was located at $P_b=15.21\pm0.02$~d, while those at 20.8~d (presumably its 2-month alias) and at 30.4~d (its first multiple) were both less likely.  The 
parameters and error bars of the best models are listed in Table~\ref{tab:pla}, while the plot associated with the 15.21-d planet model is shown in Fig.~\ref{fig:rv} 
(with a zoom on the 2021 data in Fig.~\ref{fig:rv2}).  

\begin{figure*}[ht!]
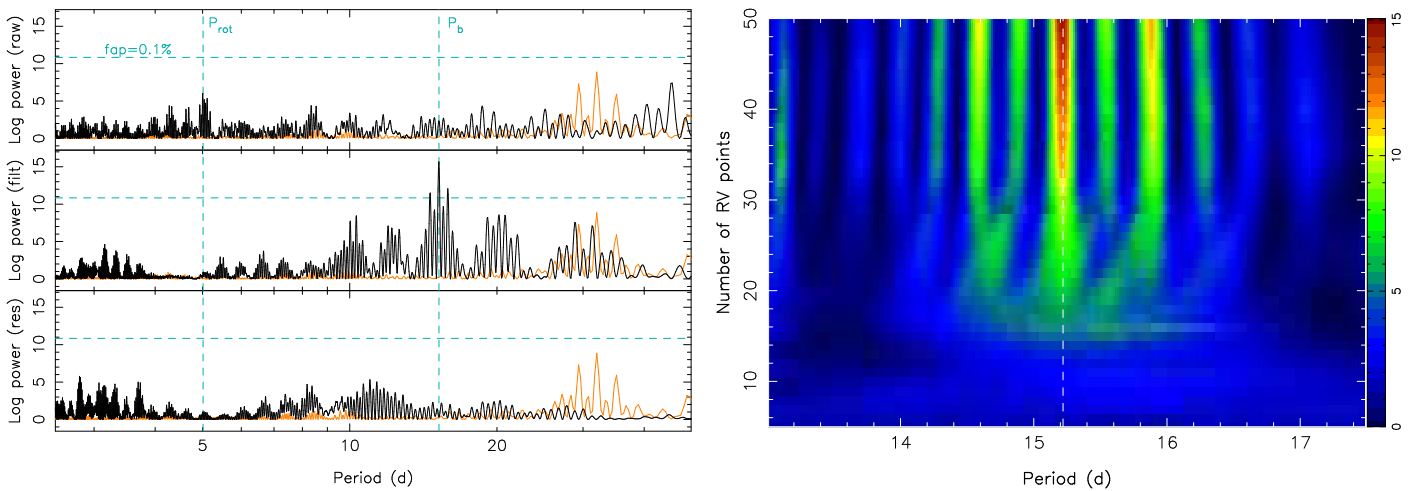

\centerline{\includegraphics[scale=0.37,angle=-90]{fig/twa7-per.ps}\hspace{3mm}\includegraphics[scale=0.32,angle=-90]{fig/twa7-stp.ps}}  
\caption[]{Left panel: periodogram of the raw (top), filtered (middle), and residual (bottom) RV data, including a {\emr tentative planet} on a 15.21-d circular orbit in the MCMC modeling.
Dashed vertical cyan lines trace the stellar rotation period and the orbital period {\emr of the tentative planet}, while the dashed horizontal line indicates a 0.1\% false-alarm probability 
\citep[\emr FAP, computed following][]{Press92} in 
the periodogram of the RV data.  No significant signal remains in the residual RVs at the main stellar and {\emr tentative planet} periods (and their aliases).  The orange curve depicts the 
periodogram of the window function. {\emr A version of this plot with frequencies on the x-axis is also provided in Fig.~\ref{fig:per2}.}   
Right panel: Stacked periodograms of the filtered RVs, as a function of the 
number of RV points included in the Fourier analysis, beginning from the first observation.  The color scale indicates the logarithmic power in the periodogram.  
The vertical dashed line traces the orbital period {\emr of the tentative planet}. } 
\label{fig:per} 
\end{figure*}

\begin{figure}[ht!]
\centerline{\includegraphics[scale=0.45,angle=-90]{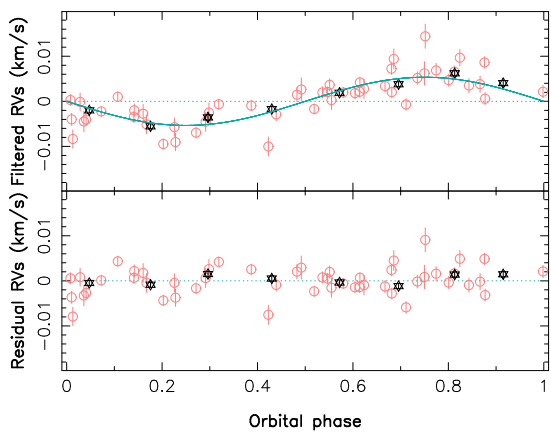}}
\caption[]{Phase-folded filtered (top) and residual (bottom) RVs for the model with a {\emr tentative planet} on a 15.21-d circular orbit.  
Red dots show the individual RV measurements with their error bars, while black stars indicate average RVs over eight even phase bins.
As in Fig.~\ref{fig:rv}, the dispersion of RV residuals is 2.8~\ms.   } 
\label{fig:rvf}
\end{figure} 

The most likely model with a planet (at an orbital period of 15.21~d) fits the RV data much better than that featuring activity only, with 1.7$\times$ smaller RV residuals and a 
logarithmic Bayes factor, $\log$~BF~=~$\Delta \log \mathcal{L}_M=11.4$ (3.4 and 2.7 higher than the models with planet at orbital periods of 20.8 and 30.4~d, respectively, see 
Table~\ref{tab:pla}).  The corresponding semi-amplitude of the planet RV signature, $K_b=5.4^{+1.3}_{-1.0}$~\ms, implies a planet minimum mass $M_b \sin i=12.5^{+3.0}_{-2.3}$~\me, 
i.e., a planet mass $M_b=0.17^{+0.04}_{-0.03}$~\mjup\ assuming an inclination of the orbital axis to the line of sight equal to that of the disk \citep[$i=13$\degr;][]{Ren21}, while the 
derived orbital period translates into an orbital distance of $a_b=0.092\pm0.006$~au.  Figure~\ref{fig:per} shows the corresponding periodograms of the raw, filtered and residual RVs, as 
well as the stacked periodogram of the filtered RVs illustrating that the 15.21-d peak strengthens as more data are added, as expected for a planet signature.  The phase-folded curve of 
the filtered RVs is shown in Fig.~\ref{fig:rvf}.  The derived planet mass is similar in the case of the 20.8-d model planet orbiting at 0.114~au, and increases to 0.28~\mjup\ for the 
30.4-d model planet at 0.147~au (see Table~\ref{tab:pla}).  

We however emphasize that $P_b$ is close to half the period of the main one-month peak in the window function (located at 32.0~d, see Fig.~\ref{fig:per}), and to 3$\times$ that of the 
stellar rotation period, flagging the putative planet 
signal as potentially spurious despite the large Bayes factor.  Moreover, the decay time of the activity jitter, $\theta_2$, now needed to be an order of magnitude shorter than the one 
derived from \Bl\ and $dT$ (see Table~\ref{tab:gpr}) so that RV variations, in particular those in 2019 and 2021, were well adjusted.  This suggested that some of the RVs might be affected 
by systematics.  Looking at how RVs vary with the barycentric Earth RV (BERV) revealed that RVs are slightly larger around BERV~$\simeq$~9~\kms.  Repeating the whole analysis without, e.g., 
the two high RV points collected on 2019 February 25 \& 26 near this BERV, reduced the Bayes factor to $\Delta \log \mathcal{L}_M=6.2$.  Last but not least, fitting the RV and $dT$ 
measurements simultaneously with a bidimensional GP to further constrain the planet parameters \citep[as advocated by][]{Rajpaul15} yielded different relative likelihoods for the planet 
models (the one with a period of 30.4~d becoming the most likely) and a marginally significant Bayes factor for the best model ($\Delta \log \mathcal{L}_M=5.5$).  For all these reasons, 
we consider this potential detection as tentative at best, until more data are available.  

Finally, we ran injection recovery tests on the LBL RVs of TWA~7 for planets with orbital periods ranging from 6 to 35~d and different conjunction epochs, and found that $K_b$ values 
of $\simeq$5~\ms\ were recovered with an average $\log$~BF~$\simeq$~5, while cases with $K_b\simeq10$~\ms\ were always detected with an average $\log$~BF of $\simeq$14. The 
$\log$~BF~$\simeq$~5 threshold translates to planet masses of 0.12, 0.16 and 0.20~\mjup\ for respective orbital periods of 7, 15 and 30~d, further confirming that the tentative signal 
reported above is very close to the detection limit of our data even though the inferred $\log$~BFs at the specific planet parameters mentioned in Table~\ref{tab:pla}) are well above the 
threshold.

\section{Activity}
\label{sec:act}

We also looked at specific lines in the SPIRou spectra of TWA~7, in particular the 1083~nm \hei\ infrared triplet (IRT) and the 1282~nm \pab\ line known to be reliable chromospheric 
activity proxies in low-mass stars.  

\begin{figure}[ht!]
\centerline{\hspace{-2mm}\includegraphics[scale=0.3,angle=-90]{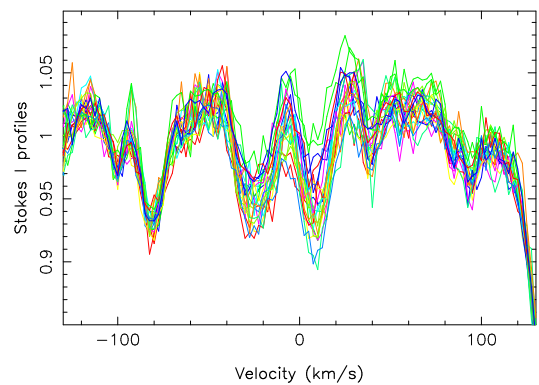}\vspace{2mm}}
\centerline{\includegraphics[scale=0.55,angle=-90]{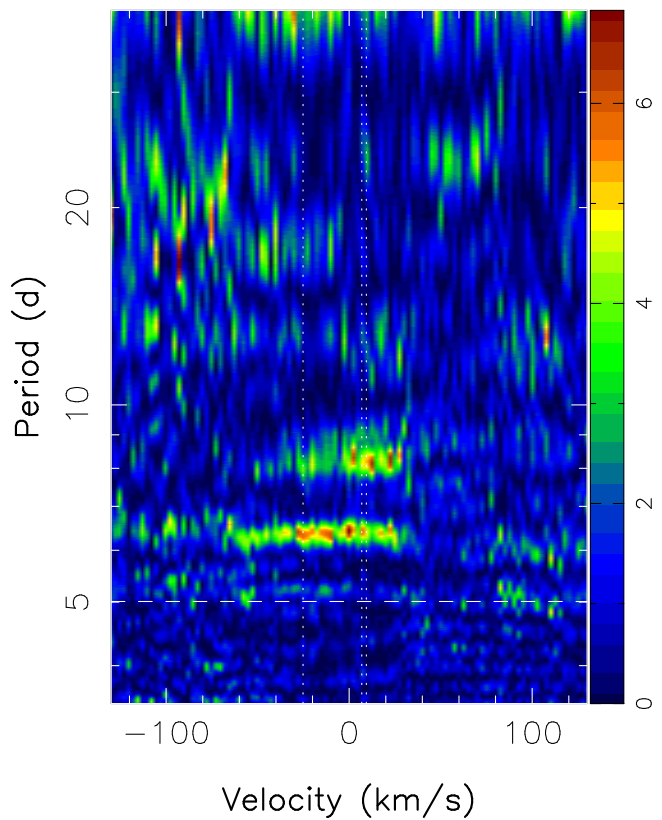}}
\caption[]{Stacked Stokes $I$ profiles (top panel) and 2D periodogram (bottom panel) of the 1083.3-nm \hei\ IRT in the stellar rest frame, for the 2021 spectra of TWA~7.  
The dashed horizontal line traces \Prot\ and the vertical dotted lines depict the velocities of the three components of the \hei\ triplet.  The color scale traces the logarithmic power in the 
periodogram.  Only the main peaks (colored yellow to red and extending over at least several velocity bins) are likely to be significant. }
\label{fig:eml}
\end{figure}

The \hei\ IRT exhibited a typical average equivalent width (EW) of $\simeq$1.6~\kms\ and only a small amount of variability at an rms level of a few \% of the continuum, with a standard 
deviation profile featuring a Gaussian shape with a full width at half maximum (FWHM) of $\simeq$30~\kms.  We saw TWA~7 flare at one epoch only (2020 May 08), with emission in the 
\hei\ core reaching an EW of 6.7~\kms\ and a FWHM of 60~\kms.  To investigate these fluctuations in more detail, we computed differential \hei\ spectra by dividing all observations by 
their median, and derived the EW changes through a Gaussian fit to the differential spectra (see Table~\ref{tab:log}), assuming a fixed centroid and FWHM (set to 0 and 50~\kms, respectively) 
as in previous works \citep[e.g.,][]{Finociety23b,Donati25}.  A periodogram of these EW changes (excluding the flare) yielded a peak at a period of 6.6~d with a false-alarm probability 
(FAP) level of 0.06\%, accompanied by a slightly weaker one at 8.3~d (presumably the 1-month alias of the main one), rather than at the stellar rotation period.  A QP GPR fit to the same 
EW changes converged on a period of $6.51\pm0.07$~d for the main peak when assuming an evolution timescale of 250~d (similar to that of \Bl\ and $dT$, see Table \ref{tab:gpr}), and 
$6.62\pm0.06$~d assuming a much longer evolution timescale (as for the periodogram).  The semi-amplitude of these EW changes was equal to $1.2\pm0.3$~\kms, with maximum emission occurring 
at a BJD of 2459298.15.  This period of 6.6~d and its 8.3~d alias were even clearly visible in the 2D periodogram of the 2021 \hei\ profiles of TWA~7 (see Fig.~\ref{fig:eml}).  

We found that \pab, with an EW of 4.0~\kms, was slightly stronger than in AU~Mic (where EW~=~3.3~\kms).  It showed much weaker temporal variability than \hei\ at a precision level of 
0.3~\kms\ rms in EW, confirming that TWA~7 is no longer accreting material from its debris disk at a detectable rate.  The flare detected in \hei\ barely showed up in \pab.  Carrying out a 
GPR fit to the \pab\ EW changes with respect to the median profile nonetheless revealed the same periodicity than with \hei\ albeit with a larger error bar ($6.52\pm0.12$~d) and a lower 
semi amplitude of fluctuations ($0.3\pm0.1$~\kms).  The \pab\ and \hei\ EW changes were reasonably well correlated given the noise ($R\simeq0.6$) and varied in phase during our observations.  
It confirmed that the detected period was real although undetected in other activity proxies.  Its origin is unclear at this stage.  

\section{Discussion}
\label{sec:dis}

We observed TWA~7 with SPIRou at CFHT, in an attempt to better characterize the star, its large- and small-scale magnetic fields and its activity pattern, and to search for potential close-in 
planets in the inner regions of its debris disk.  TWA~7 being much younger than another well studied young M dwarf, AU~Mic, hosting both a debris disk and a multi planet system \citep{Donati25}, 
makes it a key target and an interesting comparison point in the study of evolving young low-mass planet-hosting stars.  We obtained 51 Stokes $IV$ spectra of TWA~7 over 808~d covering three 
consecutive seasons (2019 to 2021), with about half of them recorded in 2021.  

We detected both the large- and small-scale magnetic fields of TWA~7.  We found that the longitudinal field \Bl\ varied from average values of $-34$~G in 2019, to $-163$~G in 2020 and $-205$~G 
in 2021.  Rotational modulation, although moderate at all epochs (semi amplitude $\simeq$50~G), was nonetheless detected clearly, with a modulation period of $5.012\pm0.007$~d, in agreement with 
literature estimates \citep[e.g.,][]{Lawson05}.  A consistent rotation period ($5.016\pm0.010$~d) was also measured from the temperature change $dT$ of TWA~7 derived by LBL, modulated with a 
semi-amplitude of $\simeq$5~K.  The small-scale magnetic field, inferred with ZeeTurbo from the Zeeman broadening of atomic lines \citep{Cristofari23}, was equal to $3.2\pm0.3$~kG, slightly 
larger than previously published measurements \citep[e.g.,][]{Yang08,Lavail19}.  

Modeling the large-scale field of TWA~7 with ZDI from LSD Stokes $IV$ profiles and $dT$ measurements simultaneously yielded a mainly poloidal and axisymmetric magnetic topology, whose main 
component was a dipole of polar strength varying from $-0.48$~kG (in 2019) to $-0.69$~kG (in 2021), tilted at an angle of $\simeq$20\degr\ to the rotation axis.  We also reconstructed a weak 
toroidal component that switched polarity between 2019 and 2020, possibly a hint of a longer-term cyclic behavior of the large-scale field.  The small-scale field inferred from the same 
modeling was equal to <$B_s$>$\simeq$2.2~kG at all epochs, smaller than the direct estimate with ZeeTurbo but consistent with the range of published values.  The absence of seasonal 
variations of <$B$> also agreed with the stationary $dT$, reported to be a reliable proxy for <$B$> \citep{Cristofari25}.  

The reconstructed field of TWA~7 is consistent with that of 
similarly young active M dwarfs \citep[e.g.,][]{Donati25b}.  It is much stronger than that derived for the same star from visible spectropolarimetric data collected in 2017 \citep{Nicholson21}, 
featuring a weak dipole and a positive polarity of the radial field in the visible hemisphere.  We suspect that this difference mostly comes from temporal variability, further suggesting that 
the large-scale field of TWA~7 undergoes long-term, possibly cyclic variations.  This difference may also partly reflect that that study was based on data collected with HARPS-Pol at visible 
wavelengths, much less sensitive to magnetic fields than SPIRou for an M dwarf.  Further observations aimed at monitoring the magnetic field of TWA~7 on a longer timescale with SPIRou 
could confirm whether the large-scale field is indeed exhibiting a Sun-like cycle with global polarity switches of its poloidal and toroidal components, and if so, would give us the 
opportunity of studying the dynamo magnetic cycle of a very young star for the first time.    

We found that RVs of TWA~7 also showed small rotational modulation (semi-amplitude $\simeq$5~\ms), and dominant fluctuations from another source with maximum power at periods of 15.2, 20.8 
or 30.4~d.  One option is that TWA~7 hosts a close-in planet in a circular orbit, most likely with an orbital period of $P_b=15.21\pm0.02$~d and a semi-amplitude of $K_b=5.4^{+1.3}_{-1.0}$~\ms, 
with a Bayes factor relative to the planet-free model large enough to qualify as a reliable detection.  Assuming an orbital inclination similar to that of the debris disk \citep[13\degr;][]{Ren21}, 
this candidate planet would correspond to a mini Saturn planet of mass $0.17^{+0.04}_{-0.03}$~\mjup\ at an orbital distance of $0.092\pm0.006$~au from TWA~7.  However, $P_b$ being close to half 
that of the one-month peak in the window function of our observations (located at 32.0~d) and to 3$\times$ the rotation period of the host star, this detection may be potentially spurious, 
despite its statistical significance.  Some of the RV fluctuations may also reflect systematics, possibly caused by residual of telluric lines, given that RVs tend to be slightly larger around 
BERV~$\simeq$9~\kms.  Besides, injection recovery tests for putative planets with orbital periods in the range 6 to 35~d and different conjunction epochs indicate that firm statistical detection 
requires in average $K_b>8$~\ms\ to obtain $\log$~BF~$>10$, confirming that the tentative RV signal reported in this paper is very close to the reliable detection limit of our data. 
For these reasons, we consider the detection of this candidate planet as only tentative, until further data are available to either validate or reject it.  If confirmed, the derived candidate 
planet would have about half the mass of that recently detected in the outer parts of the debris disk \citep{Lagrange25}.  Its existence would come as an interesting constraint for planetary 
formation models, for them to explain how a protoplanetary disk can generate both distant and close-in massive planets.  

Examining the temporal changes of the 1083~nm \hei\ IRT and 1282~nm \pab\ line of TWA~7 relative to their median profiles over the 808~d of the SPIRou observations, we found that the \hei\ IRT 
exhibited clear fluctuations at a level of a few \% of the continuum, with EWs varying with a semi-amplitude of $1.2\pm0.3$~\kms\ and a period of 6.6~d but not at the stellar rotation period.  
This period was even clearly visible in the 2D periodogram of the 2021 \hei\ spectra.  We also detected a flare at one epoch (2020 May 08) where the EW of \hei\ reached 6.7~\kms.  
On the opposite, \pab\ featured only very small EW fluctuations, at an rms level of $\simeq$0.3~\kms, confirming that TWA~7 is no longer accreting from the central regions of its debris disk.  
The 6.6~d period seen in the \hei\ IRT was nonetheless detectable in the \pab\ EW changes, in phase with \hei\ EW changes and with a typical semi-amplitude of $0.3\pm0.1$~\kms.  This period 
was not detected in the other activity proxies (\Bl\ and $dT$) nor in RVs, rendering its origin unclear.  We note that this period is equal to the beat period (of 6.60~d) between \Prot\ and 
{\emr the 20.8-d signal linked to the tentative planet} mentioned in Sec.~\ref{sec:rvs}, suggesting potential star-planet magnetospheric interactions occurring between TWA~7 and 
a putative close-in planet at a distance of 0.11~au.  Given the large-scale topology we recovered and assuming a wind from TWA~7 similar to that of AU~Mic \citep{Kavanagh21}, we find that 
the Alfven radius of TWA~7 should extend beyond 0.11~au (to ensure a sub-Alfvenic wind regime) for mass-loss rates $<$250~\msw.  Alternatively, the detected 6.6-d period could reflect the 
planet rotation period directly, although it is not clear how the planet itself could generate emission modulation in \hei\ and \pab\ lines at a level much stronger than the host star.  In 
this case, it may suggest that the planet also orbits at a period of 6.6~d, i.e., with rotation synchronized on orbital motion;  from injection recovery tests, the non-detection of this 
period in the RV SPIRou data implies an upper limit on the planet mass of $\simeq$0.12~\mjup.  

Being a younger (and slightly less massive) version of AU~Mic, TWA~7 is one of the few very best objects for studying newly born low-mass stars, their planets and debris disk at an age of 
$\simeq$10~Myr.  Its location in TWA at a distance of 33~pc makes it much more accessible than, e.g., the even younger planet-hosting low-mass star IRAS~04125+2902 
\citep{Barber24b,Donati25c}, allowing one to collect higher quality data and therefore infer more stringent constraints on planetary formation despite the low inclination of the disk 
rotation axis of TWA~7 to the line of sight.  Although we were not able to reach a definite conclusion about the presence of a close-in planet around TWA~7 from our first set of 51 
SPIRou spectra covering 808~d, we derived an upper limit on the mass of such a planet (of about half that of the recently detected distant one), and detected an enigmatic 6.6-d period 
signal in \hei\ and \pab\ lines that may relate to the existence of such a planet.  More observations are thus required to progress on this front, which will also enable us to study on a 
longer timescale the magnetic evolution of TWA~7 that may be featuring a Sun-like dynamo cycle with reversing polarities of its poloidal and toroidal components.  We thus advocate for a 
coordinated multi-instrument multi-wavelength monitoring effort of TWA~7 over the coming years to characterize in detail the star and its planetary system.

\section*{Data availability}  SPIRou data used in this study are publicly available at the Canadian Astronomy Data Center (\url{https://www.cadc-ccda.hia-iha.nrc-cnrc.gc.ca}).       

\begin{acknowledgements}
We thank an anonymous referee for carefully reading the manuscript and suggesting valuable modifications and clarifications.  
This work benefited from the SIMBAD CDS database at URL {\tt http://simbad.u-strasbg.fr/simbad} and the ADS system at URL {\tt https://ui.adsabs.harvard.edu}.  
Our study is based on data obtained at the CFHT, operated by the CNRC (Canada), INSU/CNRS (France) and the University of Hawaii.  
This project received funds from the Agence Nationale pour la Recherche (ANR, project ANR-24-CE49-3397 ORVET) and the Investissements d'Avenir program of Grenoble-Alpes University 
(project ANR-15-IDEX-02 Origin of Life).
The authors wish to recognise and acknowledge the very significant cultural role and reverence that the summit of Maunakea has always had
within the indigenous Hawaiian community.  
\end{acknowledgements}

\bibliographystyle{aa}
\bibliography{twa7}
\clearpage

\begin{appendix}

\section{SPIRou observations: additional material}
\label{sec:appA}

Table~\ref{tab:log} provides the observation log for the SPIRou spectra of TWA~7, and the measurements derived from them at each epoch.

\begin{table*}[ht!]
\caption[]{Observing log of our SPIRou observations of TWA~7}
\centering 
\resizebox{\linewidth}{!}{   
\begin{tabular}{cccccccccccc} 
\hline
BJD        & UT date & BERV   & c / $\phi$ & t$_{\rm exp}$ & S/N & $\sigma_P$            &  \Bl\  &  RV    &  $dT$ &  EW \hei &  EW \pab \\
(2459000+) &         & (\kms) &           &   (s)        & ($H$) & ($10^{-4} I_c$)      &   (G)  &  (\ms) &  (K)   & (\kms)   & (\kms)\\
\hline
-470.966103 & 14 Feb 2019 & 12.790 & 0 / 0.007 & 735.5 & 157 & 6.21 & -35.1$\pm$19.5 & -2.9$\pm$2.3 & 7.5$\pm$5.2 & 0.9$\pm$0.7 & -0.4$\pm$0.3 \\
-468.968465 & 16 Feb 2019 & 12.080 & 0 / 0.405 & 735.5 & 177 & 4.76 & -25.7$\pm$14.3 & 4.9$\pm$2.1 & -10.9$\pm$4.6 & -1.1$\pm$0.5 & -0.7$\pm$0.3 \\
-468.045981 & 17 Feb 2019 & 11.920 & 0 / 0.589 & 735.5 & 212 & 4.33 & -17.5$\pm$12.7 & -1.3$\pm$1.9 & -15.7$\pm$4.2 & -1.8$\pm$0.3 & -0.7$\pm$0.2 \\
-466.981462 & 18 Feb 2019 & 11.384 & 0 / 0.802 & 735.5 & 176 & 5.29 & -16.0$\pm$15.7 & 2.1$\pm$2.1 & -11.0$\pm$4.8 & -2.4$\pm$0.5 & -1.1$\pm$0.3 \\
-460.108062 & 25 Feb 2019 & 9.075 & 2 / 0.173 & 735.5 & 175 & 5.02 & -50.2$\pm$15.8 & 18.5$\pm$2.6 & 3.3$\pm$5.8 & -0.1$\pm$0.5 & -0.7$\pm$0.3 \\
-458.980173 & 26 Feb 2019 & 8.352 & 2 / 0.398 & 735.5 & 211 & 4.22 & -31.4$\pm$12.8 & 25.2$\pm$1.8 & -3.8$\pm$4.0 & -0.9$\pm$0.4 & -0.7$\pm$0.3 \\
-410.241521 & 16 Apr 2019 & -10.507 & 12 / 0.123 & 668.6 & 171 & 5.30 & -51.0$\pm$16.4 & -0.5$\pm$2.0 & 1.0$\pm$4.4 & -0.5$\pm$0.5 & -1.6$\pm$0.3 \\
-408.232157 & 18 Apr 2019 & -11.237 & 12 / 0.524 & 668.6 & 162 & 5.62 & -18.8$\pm$17.7 & 6.5$\pm$2.1 & -12.0$\pm$4.6 & 0.4$\pm$0.5 & -0.5$\pm$0.3 \\
-407.236126 & 19 Apr 2019 & -11.575 & 12 / 0.722 & 668.6 & 173 & 5.24 & -42.0$\pm$16.7 & -0.8$\pm$2.0 & -13.4$\pm$4.5 & -2.0$\pm$0.4 & -1.5$\pm$0.3 \\
-406.241650 & 20 Apr 2019 & -11.905 & 12 / 0.921 & 668.6 & 171 & 5.37 & -35.8$\pm$17.0 & -6.4$\pm$2.1 & -8.2$\pm$4.7 & -1.3$\pm$0.4 & -0.8$\pm$0.3 \\
-404.239260 & 22 Apr 2019 & -12.590 & 13 / 0.320 & 668.6 & 170 & 5.47 & -47.2$\pm$17.4 & -7.8$\pm$2.1 & 1.8$\pm$4.7 & 2.9$\pm$0.5 & 0.3$\pm$0.3 \\
-403.190517 & 23 Apr 2019 & -13.051 & 13 / 0.529 & 668.6 & 141 & 7.25 & -34.2$\pm$24.1 & 9.8$\pm$2.6 & -6.9$\pm$5.7 & 0.8$\pm$0.6 & -1.3$\pm$0.4 \\
-402.229905 & 24 Apr 2019 & -13.279 & 13 / 0.721 & 668.6 & 163 & 5.76 & -48.0$\pm$18.1 & 2.0$\pm$2.1 & -9.6$\pm$4.7 & -0.1$\pm$0.5 & -0.2$\pm$0.4 \\
-401.192698 & 25 Apr 2019 & -13.703 & 13 / 0.928 & 668.6 & 185 & 5.02 & -60.8$\pm$15.6 & -2.8$\pm$2.2 & -7.2$\pm$4.8 & -3.3$\pm$0.4 & -1.1$\pm$0.3 \\
-400.242973 & 26 Apr 2019 & -13.896 & 14 / 0.118 & 668.6 & 159 & 6.03 & -45.4$\pm$19.1 & 3.0$\pm$2.2 & 2.1$\pm$4.9 & -1.8$\pm$0.5 & -1.1$\pm$0.3 \\
-399.239647 & 27 Apr 2019 & -14.224 & 14 / 0.318 & 668.6 & 134 & 7.53 & 3.9$\pm$23.6 & 15.1$\pm$2.7 & 4.1$\pm$5.9 & -0.6$\pm$0.6 & -0.9$\pm$0.4 \\
-395.261370 & 01 May 2019 & -15.403 & 15 / 0.111 & 668.6 & 180 & 5.15 & -18.8$\pm$15.8 & -16.9$\pm$2.0 & 1.4$\pm$4.5 & -0.5$\pm$0.4 & -0.1$\pm$0.2 \\
\hline
-114.975903 & 05 Feb 2020 & 15.924 & 71 / 0.034 & 1203.5 & 303 & 2.78 & -192.7$\pm$8.2 & -13.0$\pm$1.4 & 3.9$\pm$3.0 & -1.1$\pm$0.2 & -0.8$\pm$0.1 \\
-22.232821 & 08 May 2020 & -17.626 & 89 / 0.539 & 1203.5 & 322 & 2.72 & -117.5$\pm$8.5 & -4.9$\pm$1.3 & 12.0$\pm$2.7 & 6.3$\pm$0.3 & 0.6$\pm$0.2 \\
-21.193862 & 09 May 2020 & -17.978 & 89 / 0.746 & 1203.5 & 305 & 2.81 & -131.7$\pm$8.7 & -5.8$\pm$1.3 & -2.5$\pm$2.8 & 1.0$\pm$0.2 & -0.1$\pm$0.1 \\
-20.240633 & 10 May 2020 & -18.113 & 89 / 0.936 & 1203.5 & 306 & 2.87 & -188.7$\pm$8.9 & -2.5$\pm$1.3 & 8.4$\pm$2.9 & 1.8$\pm$0.2 & -0.4$\pm$0.1 \\
-19.208720 & 11 May 2020 & -18.438 & 90 / 0.142 & 1203.5 & 321 & 2.70 & -201.1$\pm$8.2 & 0.6$\pm$1.4 & 14.3$\pm$3.0 & 1.0$\pm$0.2 & -0.4$\pm$0.1 \\
-18.209521 & 12 May 2020 & -18.677 & 90 / 0.341 & 1203.5 & 300 & 2.84 & -169.7$\pm$8.6 & -0.5$\pm$1.4 & 12.8$\pm$3.0 & 0.2$\pm$0.2 & -0.2$\pm$0.1 \\
-17.207107 & 13 May 2020 & -18.918 & 90 / 0.541 & 1203.5 & 216 & 4.11 & -111.4$\pm$12.5 & -2.8$\pm$1.8 & 4.2$\pm$4.0 & -0.8$\pm$0.3 & -0.2$\pm$0.2 \\
-15.202337 & 15 May 2020 & -19.385 & 90 / 0.941 & 1203.5 & 278 & 3.11 & -189.4$\pm$9.7 & -2.8$\pm$1.4 & 5.3$\pm$3.0 & 0.9$\pm$0.2 & -0.0$\pm$0.1 \\
\hline
267.009838 & 21 Feb 2021 & 10.098 & 147 / 0.249 & 1203.5 & 318 & 2.97 & -201.6$\pm$9.4 & 17.8$\pm$1.4 & 13.5$\pm$3.0 & -0.4$\pm$0.3 & 0.1$\pm$0.2 \\
267.965818 & 22 Feb 2021 & 9.832 & 147 / 0.439 & 1203.5 & 336 & 2.64 & -164.0$\pm$7.9 & 6.5$\pm$1.2 & 3.8$\pm$2.6 & -0.6$\pm$0.2 & -0.1$\pm$0.1 \\
268.981564 & 23 Feb 2021 & 9.407 & 147 / 0.642 & 1203.5 & 285 & 3.10 & -141.8$\pm$9.2 & 10.2$\pm$1.3 & -3.1$\pm$2.8 & -1.1$\pm$0.2 & -0.4$\pm$0.1 \\
271.946868 & 26 Feb 2021 & 8.337 & 148 / 0.234 & 1203.5 & 297 & 3.06 & -215.3$\pm$8.9 & 20.6$\pm$1.3 & 9.5$\pm$2.7 & 0.2$\pm$0.2 & -0.1$\pm$0.1 \\
273.994862 & 28 Feb 2021 & 7.425 & 148 / 0.642 & 1203.5 & 297 & 3.02 & -130.1$\pm$8.9 & -2.3$\pm$1.4 & -3.9$\pm$3.0 & -1.2$\pm$0.2 & -0.2$\pm$0.1 \\
275.994096 & 02 Mar 2021 & 6.637 & 149 / 0.041 & 1203.5 & 235 & 3.80 & -244.6$\pm$11.5 & 11.2$\pm$1.4 & 4.2$\pm$3.1 & -2.0$\pm$0.2 & -0.3$\pm$0.2 \\
276.918476 & 03 Mar 2021 & 6.435 & 149 / 0.226 & 1203.5 & 318 & 2.63 & -221.1$\pm$7.9 & -0.4$\pm$1.2 & 4.6$\pm$2.5 & -1.1$\pm$0.2 & -0.0$\pm$0.1 \\
277.963626 & 04 Mar 2021 & 5.913 & 149 / 0.434 & 1203.5 & 365 & 2.33 & -176.7$\pm$6.8 & -7.5$\pm$1.2 & 1.5$\pm$2.6 & 0.0$\pm$0.2 & 0.4$\pm$0.1 \\
293.899533 & 20 Mar 2021 & -0.499 & 152 / 0.614 & 1203.5 & 356 & 2.48 & -139.8$\pm$7.4 & -5.2$\pm$1.1 & -7.6$\pm$2.5 & -0.7$\pm$0.1 & 0.2$\pm$0.1 \\
294.932886 & 21 Mar 2021 & -0.999 & 152 / 0.820 & 1203.5 & 343 & 2.58 & -189.1$\pm$7.8 & -3.3$\pm$1.1 & -10.2$\pm$2.5 & -1.3$\pm$0.2 & -0.1$\pm$0.1 \\
296.941210 & 23 Mar 2021 & -1.840 & 153 / 0.221 & 1203.5 & 333 & 2.67 & -239.2$\pm$8.2 & -2.7$\pm$1.1 & 0.1$\pm$2.5 & 0.7$\pm$0.2 & 0.9$\pm$0.1 \\
297.865544 & 24 Mar 2021 & -2.048 & 153 / 0.405 & 1203.5 & 360 & 2.49 & -195.5$\pm$7.4 & -3.0$\pm$1.1 & -1.0$\pm$2.4 & 1.9$\pm$0.2 & 0.4$\pm$0.1 \\
299.868183 & 26 Mar 2021 & -2.868 & 153 / 0.805 & 1203.5 & 279 & 3.10 & -197.0$\pm$9.3 & -5.7$\pm$1.3 & -7.7$\pm$2.8 & 0.2$\pm$0.2 & 0.2$\pm$0.1 \\
300.826089 & 27 Mar 2021 & -3.160 & 153 / 0.996 & 1203.5 & 219 & 4.14 & -243.0$\pm$12.7 & 6.6$\pm$1.6 & 5.1$\pm$3.5 & -1.0$\pm$0.2 & -0.3$\pm$0.2 \\
301.874881 & 28 Mar 2021 & -3.693 & 154 / 0.205 & 1203.5 & 241 & 3.73 & -221.4$\pm$11.3 & 2.1$\pm$1.6 & 2.7$\pm$3.4 & -1.4$\pm$0.2 & -0.2$\pm$0.2 \\
304.879920 & 31 Mar 2021 & -4.906 & 154 / 0.804 & 1203.5 & 277 & 3.38 & -209.1$\pm$10.3 & -11.0$\pm$1.3 & -6.4$\pm$2.9 & 1.8$\pm$0.2 & 1.0$\pm$0.2 \\
305.890004 & 01 Apr 2021 & -5.330 & 155 / 0.006 & 1203.5 & 328 & 2.65 & -252.5$\pm$8.2 & -0.3$\pm$1.1 & -2.6$\pm$2.4 & 1.6$\pm$0.2 & 0.6$\pm$0.1 \\
326.815207 & 22 Apr 2021 & -12.911 & 159 / 0.181 & 1203.5 & 236 & 4.72 & -243.3$\pm$14.6 & 3.5$\pm$1.6 & 5.8$\pm$3.5 & -1.5$\pm$0.2 & -0.3$\pm$0.1 \\
327.789820 & 23 Apr 2021 & -13.175 & 159 / 0.375 & 1203.5 & 340 & 2.70 & -201.7$\pm$8.3 & -2.5$\pm$1.2 & 3.4$\pm$2.5 & -0.6$\pm$0.1 & -0.1$\pm$0.1 \\
328.819162 & 24 Apr 2021 & -13.578 & 159 / 0.581 & 1203.5 & 325 & 2.80 & -178.8$\pm$8.5 & -7.6$\pm$1.2 & -2.9$\pm$2.5 & -0.9$\pm$0.2 & -0.4$\pm$0.1 \\
329.836055 & 25 Apr 2021 & -13.943 & 159 / 0.784 & 1203.5 & 314 & 2.89 & -204.0$\pm$8.9 & -6.0$\pm$1.2 & -3.2$\pm$2.5 & 1.3$\pm$0.2 & 0.5$\pm$0.1 \\
331.831054 & 27 Apr 2021 & -14.562 & 160 / 0.182 & 1203.5 & 305 & 3.04 & -237.8$\pm$9.4 & 9.0$\pm$1.3 & 2.5$\pm$2.8 & -0.2$\pm$0.2 & 0.2$\pm$0.1 \\
332.811354 & 28 Apr 2021 & -14.821 & 160 / 0.377 & 1203.5 & 329 & 2.85 & -200.4$\pm$8.7 & -4.6$\pm$1.4 & 0.8$\pm$3.0 & -1.2$\pm$0.1 & -0.0$\pm$0.1 \\
334.805332 & 30 Apr 2021 & -15.413 & 160 / 0.775 & 1203.5 & 350 & 2.69 & -192.4$\pm$8.2 & -6.8$\pm$1.2 & -5.0$\pm$2.6 & -1.1$\pm$0.2 & -0.2$\pm$0.1 \\
335.786838 & 01 May 2021 & -15.664 & 160 / 0.971 & 1203.5 & 336 & 2.70 & -234.1$\pm$8.2 & -0.4$\pm$1.2 & -0.4$\pm$2.5 & 1.7$\pm$0.2 & 0.7$\pm$0.1 \\
336.826623 & 02 May 2021 & -16.057 & 161 / 0.178 & 1203.5 & 291 & 3.12 & -248.0$\pm$9.5 & -1.0$\pm$1.2 & 3.4$\pm$2.7 & 0.0$\pm$0.1 & -0.0$\pm$0.1 \\
\hline
\end{tabular}}  
\tablefoot{For each visit, we list the barycentric Julian date BJD, the UT date, the barycentric Earth RV (BERV), the rotation cycle c and phase $\phi$ (computed as indicated in Sec.~\ref{sec:obs}),
the total observing time t$_{\rm exp}$, the peak S/N in the spectrum (in the $H$ band) per 2.3~\kms\ pixel, the noise level in the LSD Stokes $V$ profile,
the estimated \Bl, LBL RV, $dT$ and the \hei\ and \pab\ EW change with corresponding error bars. }
\label{tab:log}
\end{table*} 

\begin{figure*}
\centerline{\includegraphics[scale=0.55,bb=20 40 900 610]{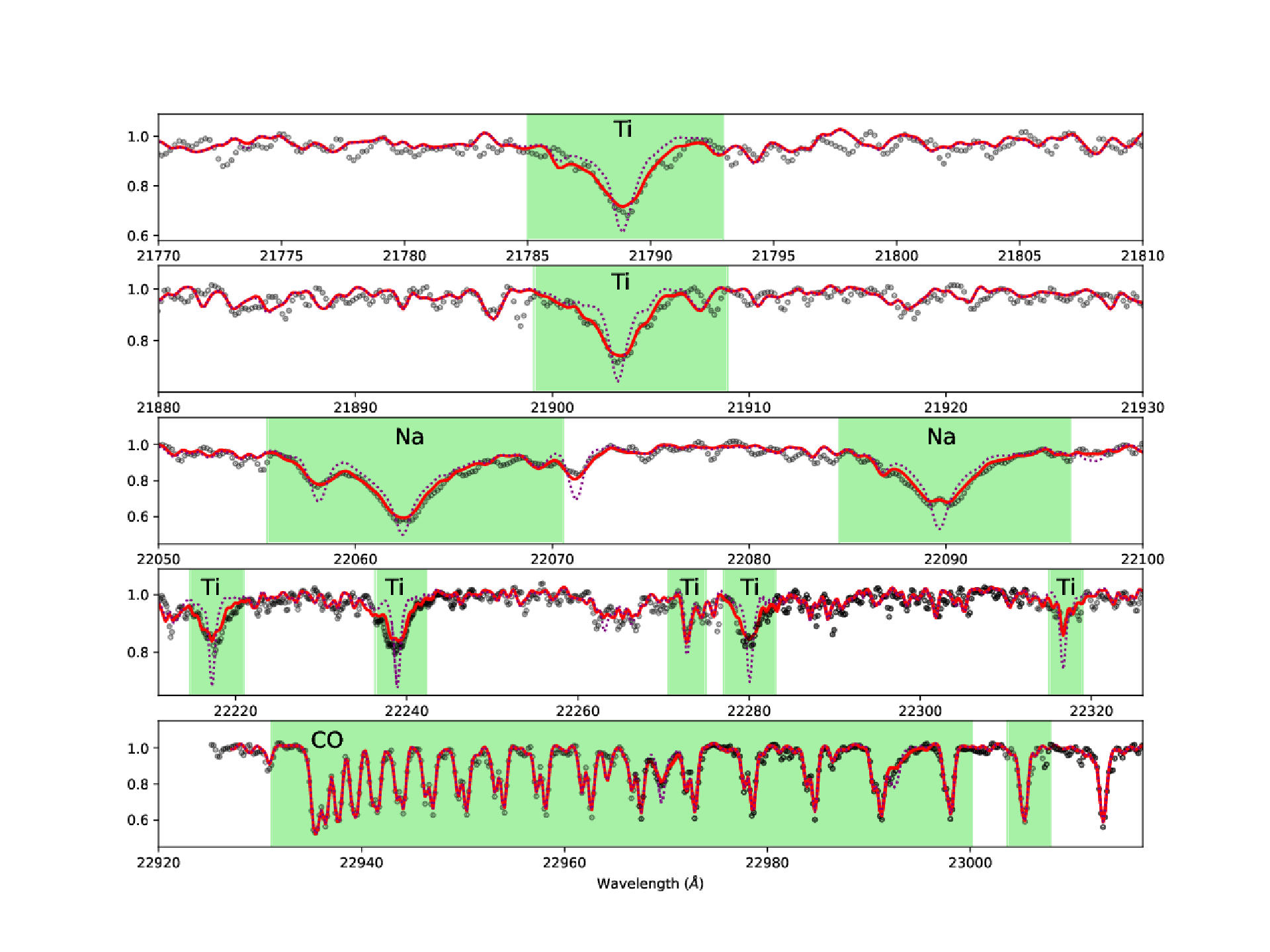}}
\caption[]{Small portion of our template SPIRou spectrum of TWA~7 in the $K$ band (black circles), along with the optimal fit achieved
with ZeeTurbo (red line) using the modeling approach of \citet{Cristofari23}.  The green areas indicate spectral regions considered in the
fit, and the dotted line shows the model with no magnetic field. }
\label{fig:spc}
\end{figure*}

\section{Magnetic field \& temperature changes: additional material}
\label{sec:appB}

Figure~\ref{fig:gpb2} shows the \Bl, and $dT$ curves with corresponding GPR fits, zooming on the 2021 data.  

\begin{figure*}[ht!]
\centerline{\includegraphics[scale=0.39,angle=-90]{fig/twa7-gpb21.ps}\vspace{2mm}}
\centerline{\includegraphics[scale=0.39,angle=-90]{fig/twa7-gpt21.ps}}
\caption[]{Same as Fig.~\ref{fig:gpb}, zooming on the 2021 data.} 
\label{fig:gpb2}
\end{figure*}

\section{ZDI modeling: additional material}
\label{sec:appC}

\begin{figure*}[ht!]
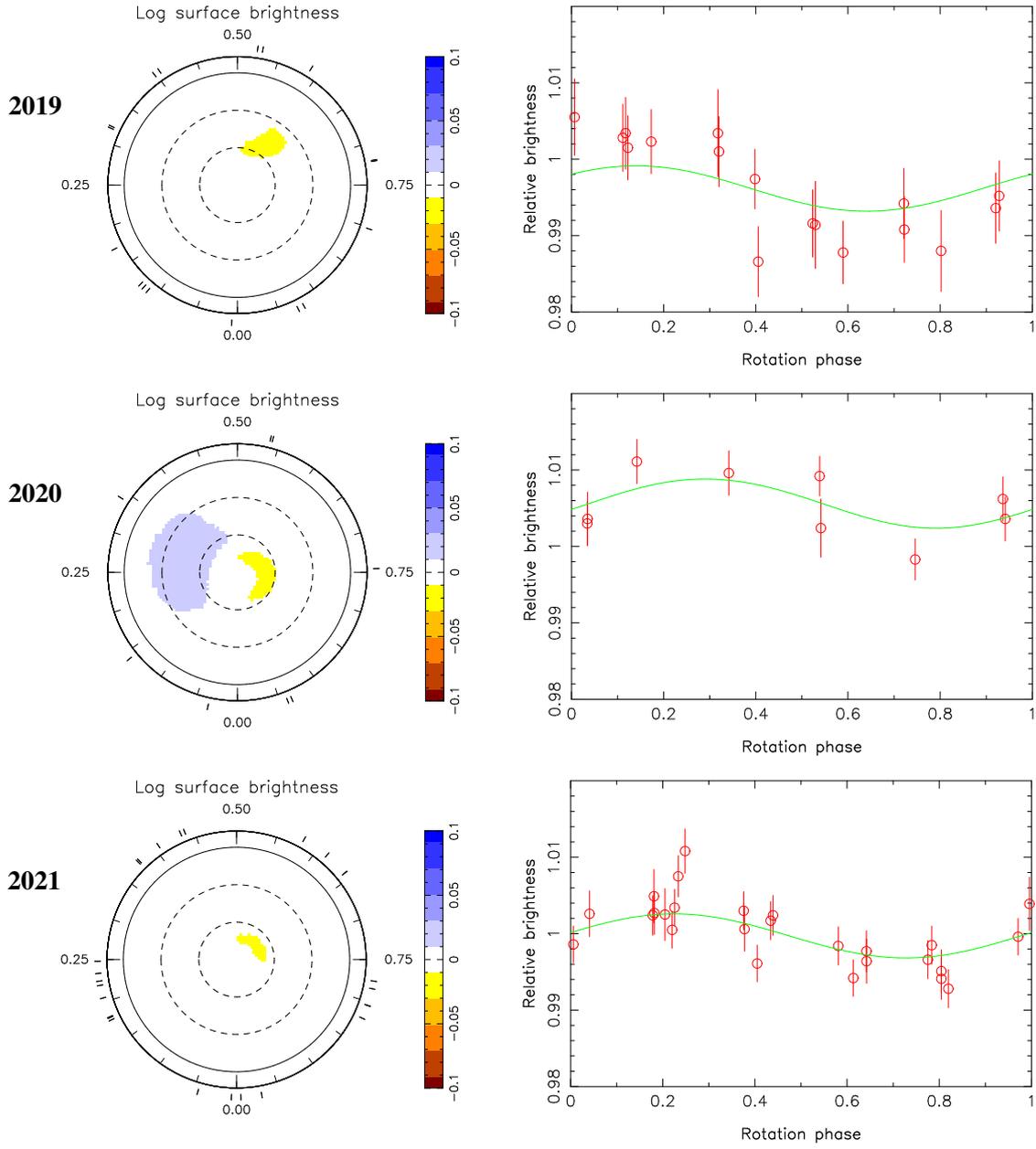

\centerline{\large\bf 2019\raisebox{0.3\totalheight}{\includegraphics[scale=0.18,angle=-90]{fig/twa7-mapi19.ps}\hspace{9mm}\includegraphics[scale=0.32,angle=-90]{fig/twa7-ph19.ps}}\vspace{3mm}}
\centerline{\large\bf 2020\raisebox{0.3\totalheight}{\includegraphics[scale=0.18,angle=-90]{fig/twa7-mapi20.ps}\hspace{9mm}\includegraphics[scale=0.32,angle=-90]{fig/twa7-ph20.ps}}\vspace{3mm}}
\centerline{\large\bf 2021\raisebox{0.3\totalheight}{\includegraphics[scale=0.18,angle=-90]{fig/twa7-mapi21.ps}\hspace{9mm}\includegraphics[scale=0.32,angle=-90]{fig/twa7-ph21.ps}}}
\caption[]{Brightness maps (left panels) reconstructed simultaneously with the magnetic maps of Fig.~\ref{fig:map} with ZDI, and photometric light curves inferred from $dT$ estimates (red circles, 
right panels) with ZDI fits (green curves).  In the maps, yellow and blue depict regions darker and brighter than the quiet photosphere, respectively.  }
\label{fig:pho}
\end{figure*}

Figure~\ref{fig:pho} shows, for each epoch, the ZDI fits to the photometric curves derived from $dT$, and the brightness maps derived simultaneously with the magnetic maps of Fig.~\ref{fig:map}.  

\section{RV modeling: additional material}
\label{sec:appD}

\begin{figure}[ht!]
\centerline{\includegraphics[scale=0.37,angle=-90]{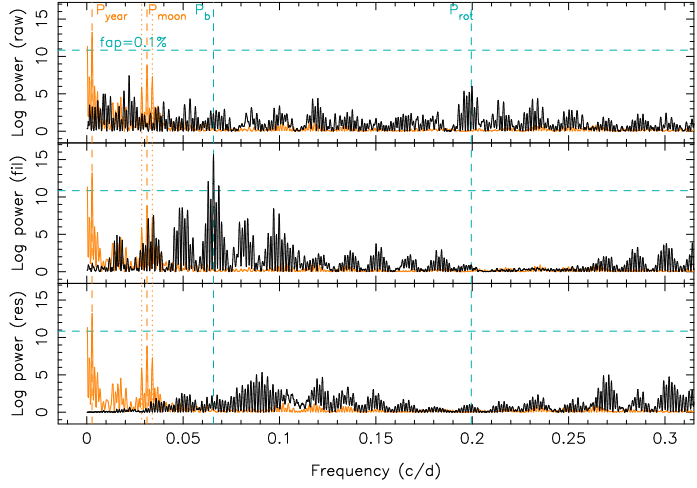}}  
\caption[]{\emr Same as the left panel of Fig.~\ref{fig:per} with frequencies (instead of periods) on the x-axis.  The orange dashed and dotted vertical lines depict the 1-yr 
frequency $P_{\rm year}$ and the peak of the window function (at 32.0~d, with its 1-yr aliases) roughly corresponding to the lunar frequency $P_{\rm moon}$. } 
\label{fig:per2} 
\end{figure}

Figure~\ref{fig:rv2} shows the RV curve and MCMC + GPR fit, zooming on the 2021 data, {\emr while Fig.~\ref{fig:per2} shows the periodogram of Fig.~\ref{fig:per} (left panel) 
with frequencies on the x-axis.}   

\begin{figure}[ht!]
\centerline{\includegraphics[scale=0.39,angle=-90]{fig/twa7-rv21.ps}}
\caption[]{Same as Fig.~\ref{fig:rv}, zooming on the 2021 data}  
\label{fig:rv2}
\end{figure}

\section{Activity: additional material}
\label{sec:appE}

\begin{figure}[ht!]
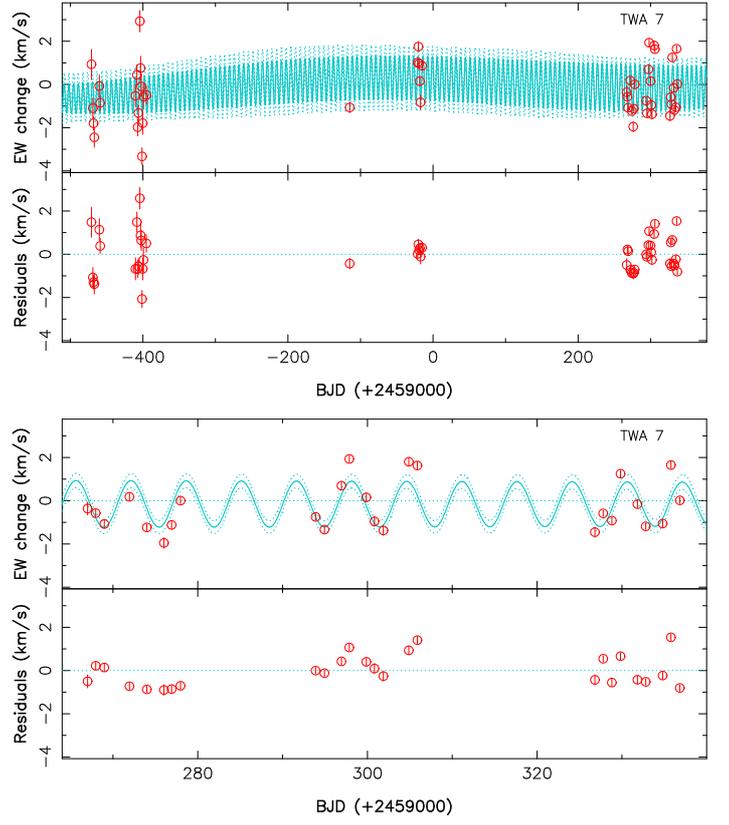

\centerline{\includegraphics[scale=0.39,angle=-90]{fig/twa7-ew.ps}\vspace{2mm}}
\centerline{\includegraphics[scale=0.39,angle=-90]{fig/twa7-ew21.ps}}
\caption[]{Same as Fig.~\ref{fig:gpb} for the EW of the differential \hei\ spectra (top panel) with a zoom on the 2021 data (bottom panel), assuming $\theta_3=250$~d.  
The data point recorded when TWA~7 flared was removed from this analysis.  } 
\label{fig:ew}
\end{figure}

Figure~\ref{fig:ew} shows the EW of the differential \hei\ spectra fitted with GPR, with a zoom on the 2021 data.  

\FloatBarrier 
\clearpage

\end{appendix}
\end{document}